\documentclass[]{aastex631}
\usepackage{amsmath}
\usepackage{multirow}
\usepackage{booktabs}
\usepackage{hyperref}
\begin{document}


\title{CU-JADE: A Method for Traversing Extinction Jumps along the Line of Sight}

\author[0009-0002-2379-4395]{Shiyu Zhang}
\affiliation{Purple Mountain Observatory and Key Laboratory of \\
Radio Astronomy, Chinese Academy of Sciences, Nanjing 210034, \\
People’s Republic of China}
\affiliation{School of Astronomy and Space Science, University of \\
Science and Technology of China, 96 Jinzhai Road, Hefei 230026, \\
People’s Republic of China}

\author[0000-0002-0197-470X]{Yang Su$^\dagger$}
\affiliation{Purple Mountain Observatory and Key Laboratory of \\
Radio Astronomy, Chinese Academy of Sciences, Nanjing 210034, \\
People’s Republic of China}
\affiliation{School of Astronomy and Space Science, University of \\
Science and Technology of China, 96 Jinzhai Road, Hefei 230026, \\
People’s Republic of China}
\author[0000-0003-3151-8964]{Xuepeng Chen}
\affiliation{Purple Mountain Observatory and Key Laboratory of \\
Radio Astronomy, Chinese Academy of Sciences, Nanjing 210034, \\
People’s Republic of China}
\affiliation{School of Astronomy and Space Science, University of \\
Science and Technology of China, 96 Jinzhai Road, Hefei 230026, \\
People’s Republic of China}

\author[0000-0001-8060-1321]{Min Fang}
\affiliation{Purple Mountain Observatory and Key Laboratory of \\
Radio Astronomy, Chinese Academy of Sciences, Nanjing 210034, \\
People’s Republic of China}
\affiliation{School of Astronomy and Space Science, University of \\
Science and Technology of China, 96 Jinzhai Road, Hefei 230026, \\
People’s Republic of China}

\author[0000-0002-7489-0179]{Fujun Du}
\affiliation{Purple Mountain Observatory and Key Laboratory of \\
Radio Astronomy, Chinese Academy of Sciences, Nanjing 210034, \\
People’s Republic of China}
\affiliation{School of Astronomy and Space Science, University of \\
Science and Technology of China, 96 Jinzhai Road, Hefei 230026, \\
People’s Republic of China}

\author[0000-0003-2549-7247]{Shaobo Zhang}
\affiliation{Purple Mountain Observatory and Key Laboratory of \\
Radio Astronomy, Chinese Academy of Sciences, Nanjing 210034, \\
People’s Republic of China}

\author[0000-0003-4586-7751]{Qing-Zeng Yan}
\affiliation{Purple Mountain Observatory and Key Laboratory of \\
Radio Astronomy, Chinese Academy of Sciences, Nanjing 210034, \\
People’s Republic of China}

\author[0000-0002-0409-7466]{Xin Liu}
\affiliation{Purple Mountain Observatory and Key Laboratory of \\
Radio Astronomy, Chinese Academy of Sciences, Nanjing 210034, \\
People’s Republic of China}
\affiliation{School of Astronomy and Space Science, University of \\
Science and Technology of China, 96 Jinzhai Road, Hefei 230026, \\
People’s Republic of China}

\author[0000-0002-6388-649X]{Miaomiao Zhang}
\affiliation{Purple Mountain Observatory and Key Laboratory of \\
Radio Astronomy, Chinese Academy of Sciences, Nanjing 210034, \\
People’s Republic of China}

\author[0000-0002-3904-1622]{Yan Sun}
\affiliation{Purple Mountain Observatory and Key Laboratory of \\
Radio Astronomy, Chinese Academy of Sciences, Nanjing 210034, \\
People’s Republic of China}
\affiliation{School of Astronomy and Space Science, University of \\
Science and Technology of China, 96 Jinzhai Road, Hefei 230026, \\
People’s Republic of China}

\author[0000-0001-7768-7320]{Ji Yang}
\affiliation{Purple Mountain Observatory and Key Laboratory of \\
Radio Astronomy, Chinese Academy of Sciences, Nanjing 210034, \\
People’s Republic of China}

\correspondingauthor{Yang Su}
\email{yangsu@pmo.ac.cn}



\begin{abstract}

Although interstellar dust extinction serves as a powerful distance estimator, the solar system’s location within the Galactic plane complicates distance determinations, especially for molecular clouds (MCs) at varying distances along the line of sight (LoS). The presence of complex extinction patterns along the LoS introduces degeneracies, resulting in less accurate distance measurements to overlapping MCs in crowded regions of the Galactic plane. 
In this study, we develop the CUSUM-based Jump-point Analysis for Distance Estimation (CU-JADE), a novel method designed to help mitigate these observational challenges. The key strengths of CU-JADE include: (1) sensitivity to detect abrupt jumps in Distance-$A_{\lambda}$ ($D$-$A$) datasets, (2) minimal systematic errors as demonstrated on both mock and observed data, and (3) the ability to combine CUSUM analysis with multiwavelength data to improve the completeness of distance measurements for nearby gas structures, even for extinction values as low as $\Delta A_{V} \gtrsim 0.15$ mag. By combining CO survey data with a large sample of stars characterized by high-precision parallaxes and extinctions, we uncovered the multilayered molecular gas distribution in the high-latitude Cepheus region. We also determined accurate distances to MCs beyond the Cygnus Rift by analyzing the intricate structure of gas and extinction within the Galactic plane. Additionally, we constructed a full-sky 3D extinction map extending to 4 kpc, which provides critical insights into dense interstellar medium components dominated by molecular hydrogen. These results advance our understanding of the spatial distribution and physical properties of MCs across the Milky Way. 
\end{abstract}

\keywords{Interstellar medium (847) --- Interstellar extinction (841) --- Distance measure (395) --- Molecular clouds (1072)}

\section{Introduction} \label{sec:intro}
Molecular clouds (MCs), as the dense part of the interstellar medium (ISM), are the sites of star formation and evolution \citep{2001ApJ...547..792D, 2010ApJ...724..687L}. Molecular gas also plays a crucial role in the transformation between different phases of interstellar matter, as well as in characterizing the structure and evolution of galaxies \citep{2015ARA&A..53..583H}. Therefore, determining the distances to MCs is essential for understanding the distribution and properties of molecular gas, as well as the formation mechanisms of stars within these dense clouds in the Milky Way. Since molecular hydrogen ($\rm H_{2}$) does not exhibit significant radiation at radio frequencies, CO, which is the second most abundant molecule after hydrogen, serves as an excellent indicator for studying the distribution of molecular gas and other fundamental physical properties \citep{1982ApJ...262..590F, 1987ApJ...319..730S, 2008ApJ...679..481P, 2009ApJ...692...91G, 2013ARA&A..51..207B, 2023ASPC..534....1C}.

The selective absorption of starlight by the interstellar dust and gas leads to the dimming and reddening of starlight \citep{1982ApJ...262..590F, 1998ApJ...500..525S}. Utilizing the large star samples with high precision parallaxes and extinctions, we can investigate the characteristics of dust absorption in the infrared and visible spectrum. MCs traced by CO are typically accompanied by high extinction (mixed with a significant quantity of dust grains), yet the presence of dust extinction is not always correlated with CO-bright MCs \citep{1983A&A...128..212M, 2001ApJ...547..792D}. 
The dust distribution exhibits a wide dynamic range from very dense cores to nearly transparent regions \citep{2003ARA&A..41..241D, 2008ApJ...679..481P, 2011ApJ...737...12L}. In addition, in low column density regions, significant dust emission can be observed in the mid-infrared, as well as the reddening effects in the optical and ultraviolet \citep{2011piim.book.....D, 2020A&A...641A..12P, 2021ApJ...911...55S}. 
Although CO emission is very weak or even absent in low column density regions, $\rm H_{2}$ may exist based on observations and models \citep{2005Sci...307.1292G, 2010ApJ...716.1191W, 2020A&A...639A..26K, 2024A&A...685L..12L, 2024A&A...682A.161S}. We thus may use the extinction features to trace CO-faint or even CO-dark MCs. 

The detailed 3D distribution and kinematic properties resolved along the line of sight (LoS) are essential for the study of MCs.
For example, the superposition of crowded molecular gas components along the LoS directly impacts the mass-size relation of MCs \citep[e.g.,][]{2019MNRAS.490.2648B, 2020ApJ...898....3L}, leading to misinterpretation of the indices derived from various statistical samples. 
By examining a sample of stars within a specific distance range, some studies show that the polarization patterns of the ISM do not closely match the orientation of the magnetic field, which may indicate that the observations are affected by the superposition of multiple gas layers \citep{2020A&A...643A.151R, 2024A&A...684A.162P, 2024arXiv240603765P}. 

Indeed, MCs near the Galactic disk often exhibit multilayer distributions \citep[e.g., toward the Aquila and Cygnus region;][]{2020ApJ...893...91S, 2024AJ....167..220Z}. This is an inevitable result of the superposition of the clouds at different distances along the LoS, leading to difficulties in determining the exact distances of MCs at different layers, especially in the Galactic plane. The extinction-based methods map the distributions of MCs by analyzing stellar spatial distributions in the ISM and identifying distance-dependent extinction features along the LoS. This approach was pioneered by \citet{1923AN....219..109W} and \citet{1937dss..book.....B}. Over the past several years, the Gaia mission has revolutionized the field with its high-precision parallax measurements, catalyzing a rapid growth in the development of 3D extinction maps within the solar neighborhood \citep{2019MNRAS.483.4277C, 2019ApJ...887...93G, 2019A&A...625A.135L, 2022A&A...664A.174V, 2024A&A...685A..82E}. While these maps exhibit consensus on large-scale features, notable discrepancies emerge in complex environments. 

Meanwhile, extinction-jump methods, which employ single-extinction-jump models, are often applicable to high-latitude regions \citep{2014ApJ...786...29S, 2019A&A...624A...6Y, 2019ApJ...879..125Z, 2020ApJ...891..137Z, 2022MNRAS.511.2302G}. Based on the identification of the MCs' silhouette, which highlights the target MCs with selection of on-cloud stars and field stars, some works investigate the distance measurements for clouds near Galactic plane \citep{2019ApJ...885...19Y, 2020MNRAS.493..351C, 2024AJ....167..220Z}. However, the identification of MCs depends on the algorithms and changes with it. On the other hand, in the inner Galactic region, the crowded MCs along the LoS cause significant extinction, leading to difficulties in detecting jumps in some more distant MCs and/or MCs with weak emission. 

Aiming to detect small extinction jumps along the LoS, we developed a data-driven method that minimizes dependence on complex modeling while
remaining sensitive enough to capture extinction jumps caused by relatively high column density gas. 
By analyzing statistically these robust extinction jumps along the LoS, our approach facilitates the reconstruction of dust characteristics across the entire sky, which exhibit a stronger correlation with the molecular gas environment. We designate this new method as the CUSUM-based Jump-point Analysis for Distance Estimation (CU-JADE). 

This paper is organized as follows. In Section \ref{sec2}, we introduce the data we used in this work. Then, we describe how the CU-JADE method works based on the characteristics of the LoS stellar parameters in Section \ref{sec3}, followed by validation of the method using mock and observed data in Section \ref{sec4}. In Section \ref{sec5}, we further implement the method on multilayer gas distance measurements for two typical regions toward the Cepheus and the Cygnus. We also present a 3D dust map that extends to $\sim 4$ kpc for whole sky. Finally, in Section \ref{sec6}, we summarize the main results and conclusions.

\section{Data} \label{sec2}

\subsection{Molecular Gas Data}
We utilize various CO surveys to trace the distribution of molecular gas. The spatial distribution, structural characteristics, and kinematics of these MCs will assist in selecting suitable stellar samples for measuring various distance components and investigating their correlation with  corresponding extinction features. 
Considering the different coverage and resolution of CO surveys in the Galactic plane and at high latitudes, 
we have adopted different datasets to measure and analyze the MC distances. 
\subsubsection{MWISP Data} \label{mwispcyg}
The CO data in the Galactic plane originates from the Milky Way Imaging Scroll Painting \citep[MWISP project, for more details, refer to][]{2019ApJS..240....9S}. In brief, the spatial resolution of the observational data is $\sim50''$, and the spectral resolution is $\sim0.2$ $\rm km~s^{-1}$. The processed 3D data cubes, referred to as position-position-velocity space, have a grid spacing of $30''$. The typical root mean square (rms) noise level is $\sim0.5$ K for $\rm ^{12}CO$ and $\sim0.3$ K for $\rm ^{13}CO$ and $\rm C^{18}O$. In this work, we focus on MCs located in a specific region of the Galactic plane toward the Cygnus region \citep[i.e., $72^{\circ} \leq l \leq 87^{\circ}$,  $|b| \leq 5^{\circ}.1$;][]{2024AJ....167..220Z}. 
\subsubsection{1.2 m CfA Data}
The 1.2 m CfA survey \citep{2001ApJ...547..792D} incorporates data from 37 individual $\rm ^{12}CO$ ($J$ = 1--0) line surveys covering most of the Galactic plane. With an angular resolution of $8'.6$, the surveys provide exceptional data homogeneity. Notably, it offers velocity coverage of $\pm$ 47.1 $\rm km~s^{-1}$ and significantly improved sensitivity of 0.18 K in 0.65 $\rm km~s^{-1}$ channels for high-latitude northern sky \citep{2022ApJS..262....5D} compared to the MWISP survey. We utilized the CfA dataset to systematically investigate dust-gas correlations in the Cepheus region, particularly focusing on the relationship between CO structures and the corresponding extinction features in $100^{\circ} \leq l \leq 120^{\circ}$, and $0^{\circ} \leq b \leq20^{\circ}$. 
\subsection{Stellar Parameters}
We use stellar parameters (e.g., extinction) and parallaxes from different star samples to measure distances of extinction structures. In this work, we generate 3D extinction maps using different stellar catalogs (Section \ref{gaiadr3} for Gaia DR3, Section \ref{starhorse} for SHEDR3, and Section \ref{zgr23} for ZGR23). Our 3D extinction maps are broadly similar to known studies \citep[e.g.,][]{2022A&A...664A.174V}. However, we did not conduct further quantitative comparison of the systematic differences between the catalogs. 
\subsubsection{Gaia DR3}
\label{gaiadr3}
Gaia's Data Release 3 \citep[i.e., Gaia DR3;][]{2023A&A...674A...1G, 2023A&A...674A..27A} used the General Stellar Parameterizer from Photometry (GSP-Phot) to fit the BP/RP spectrum, parallax, and apparent $G$ magnitude simultaneously with a Bayesian forward-modelling approach. Over 470 million sources with $G<19$ were released. The $G$-band ($\lambda$ centered at 673 nm) extinction $A_{G}$ was constrained to be positive with roughly an error $\sim0.06$ mag. For the Cygnus region near the Galactic plane, we retrieved the same stellar sample from Gaia DR3 and performed a comparative analysis in line with previous results \citep{2024AJ....167..220Z}. Additionally, the statistical mock data on parallaxes and extinction discussed in Section \ref{sec4} are also derived from this sample. 

\subsubsection{SHEDR3}
\label{starhorse}
StarHorse EDR3 \citep[in the following, SHEDR3;][]{2022A&A...658A..91A} had released a catalog featuring 362 million star samples, which derive distances $d$, $V$-band extinctions (at $\lambda$ = 542 nm) $A_{V}$, ages $\tau$, masses $m_{*}$, effective temperatures $T_{\rm eff}$, etc. These stars, whose parameters are derived from the data of Gaia Early Data Release 3 \citep[Gaia EDR3;][]{2021A&A...649A...1G}, have been cross-matched with the photometric catalogs of Pan-STARRS1, SkyMapper, 2MASS, and AllWISE. The high precision of the Gaia EDR3 data, combined with the extensive wavelength coverage of the additional photometric surveys and the new stellar-density priors incorporated in the improved StarHorse code \citep{2018MNRAS.476.2556Q}, reveals detailed substructures in the Milky Way and beyond. We take all stars in SHEDR3 sample with relative distance uncertainties smaller than 0.2. The derived distances $d$ and extinction $A_{V}$ are used to construct the all-sky dust map with deeper extinction by CU-JADE.  

\subsubsection{ZGR23}
\label{zgr23}
To effectively compare different extinction estimation methodologies (see details in Section \ref{zxymap}), we adopted the comprehensive stellar catalog (hereafter ZGR23) published by \citet{2023MNRAS.524.1855Z}, which provides astrophysical parameters for 220 million stars. This catalog was constructed using a forward-modeling framework, which was trained on a carefully curated subset of high-resolution stellar spectra from the Large Sky Area Multi-Object Fiber Spectroscopic Telescope, featuring a spectral resolution of $R\sim1800$. The construction process also integrated multiwavelength datasets, including Gaia BP/RP low-resolution spectra ($R\sim$ 30--100), 2MASS near-infrared photometry ($J$, $H$, $K$s bands), and unWISE mid-infrared photometry ($W$1, $W$2 bands) \citep{2023A&A...674A...1G, 2023A&A...674A...3M, 2019ApJS..240...30S, 2006AJ....131.1163S, 2010AJ....140.1868W}. The derived parameters encompass critical astrophysical quantities, such as distance estimates ($1/\varpi$; parallax-based), interstellar extinctions ($A_V$), and fundamental stellar properties, $\Theta \equiv (\text{T}_{\text{\rm eff}}, ~\log g, ~[\text{Fe}/\text{H}])$. 

To generate the extinction map, we implemented additional section criteria to ensure data reliability and accuracy:
1. Quality Control: Sources were included only if their $\text{quality\_flags}<$ 8;
2. Distance Constraint: A distance cutoff of $1/\varpi<$ 5 kpc was applied;
3. Precision Threshold: A relative parallax precision threshold of $\sigma_{\varpi}/\varpi<$ 0.2 was imposed to minimize uncertainties in distance measurements.
These criteria resulted in a refined sample of 123,284,327 stars, enabling accurate 3D extinction mapping and systematic comparison of different methods. 

\subsection{Masers}
To verify the accuracy of our method, we need MCs with precise distance measurements. Distances derived from observations of masers using Very Long Baseline Interferometry \citep[][]{2006Sci...311...54X, 2014ApJ...783..130R} are considered to be among the most reliable measurements available. Class II masers \citep[e.g.,][]{1986ApJ...308..592C} can effectively trace the dense molecular gas clumps in high-mass star-forming regions (HMSFRs). In particular, many works \citep[e.g.,][]{2019ApJ...885..131R, 2020PASJ...72...50V, 2023ApJ...953...21H} have measured the LoS velocities, proper motions, and trigonometric parallaxes of water and methanol masers associated with over 200 HMSFRs. 
We collected 75 masers with accurate parallax measurements within 3 kpc for validation in Section \ref{validatemaser}. 
 

\subsection{Young Open Clusters}
We utilize young open clusters \citep[YOCs;][]{2023A&A...673A.114H} as tracers of star-forming regions. We select clusters with more than 30 stars and with distance uncertainty $<5$\%. The age of the star cluster is limited to $log~Age~(yr)\lesssim$ 7 in order to establish the connection between the star-forming region and the corresponding gas-rich environment. That is, these YOCs might not move far from their birthplace, representing the dense molecular gas therein. 

\begin{figure}[ht!]
\centering
\includegraphics[width=18cm,angle=0]
{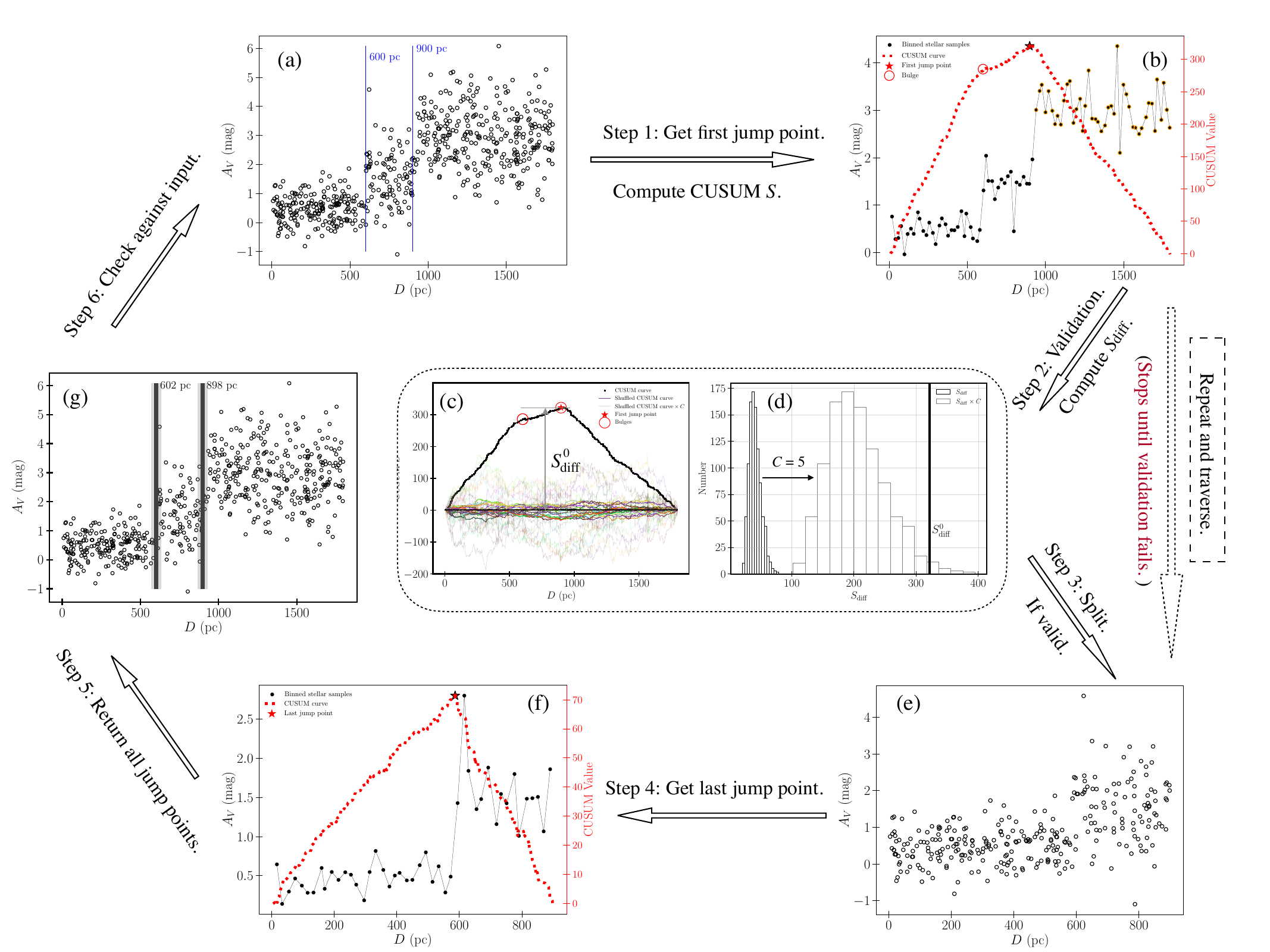}
\caption{This flowchart illustrates the process of identifying extinction-jump points in a $D$-$A$ diagram using the CU-JADE method. The process is demonstrated through a simple example with two jump points. Initialization: Begin with a set of simulated stellar samples exhibiting two jump points on the $D$-$A$ diagram (subplot (a)). These jump points are indicated by blue lines.
CUSUM Statistic Calculation: Compute the CUSUM statistic $S$ to locate the first jump point (Step 1). Subplot b shows the CUSUM curve, where black points represent binned stellar samples at intervals of 20 pc, and the red curve represents the calculated CUSUM values using Equation (3). The position with the maximum $S$ value, marked by a red star, identifies the first jump point. Validity check: Determine if the first jump point is valid (Step 2). This involves calculating the $S_{\rm diff}$ value and comparing it to the $S_{\rm diff}$ background (see Section 3.2 for details). Subplots (c) and (d) illustrate this comparison using percentile, similar to a confidence level, with a parameter $C$ for adjustment. Splitting and recalculation: If the first jump point is confirmed, split the $D$-$A$ data at this position (Step 3, see subplots (b) and (e)). Recalculate the separate datasets and repeat the process, as indicated by the dashed arrow. This begins the traversal of all extinction-jump points (suggested by bulges on the CUSUM curve). Final jump point identification: Since only two jump points were initially set, the next round of calculations locates the last jump point (Step 4, subplot (f)). And the process enables us to return all detected jump point positions (Step 5). Validation: Compare and check the detected jump point positions against the preset positions (Step 6, subplots (g) and (a)).
\label{fig1}}
\end{figure}

\section{Methods} \label{sec3}

\subsection{The CUSUM Chart-based method} \label{subsec:31}
Change point detection is the task of identifying changes in a underlying model of a signal or time series \citep{TRUONG2020107299}, which is widely implemented in various fields such as signal processing,
climate science, economics, finance, medicine, speech/image analysis, and bioinformatics  \citep{Aminikhanghahi2017, li2024fastcpd}. The well-established methods for fast detection include Bayesian analysis \citep{b9b00a23}, cumulative sum control chart \citep[CUSUM Chart;][]{Riaz2011, alanqary2021change, fearnhead2024}, sequential updating \citep{pmlr-v206-zhang23b}, and pruned exact linear time \citep[PELT;][]{1011648GDW}, etc. Bayesian analysis, dynamic programming, and PELT are characterized by cost function and a constraint on the number of changes. 
The implementation of these methods can be significantly labor-intensive when applied to long data series, due to the iterative calculation of the cost value across various segments of the data \citep{li2024fastcpd}.
CUSUM is a kind of control chart designed to detect deviations from a target value and monitor the mean of a process. The concept of CUSUM is remarkably straightforward, making it a versatile tool for various applications. 

The increment of LoS extinctions is statistically discernible as a multiple jumps in Gaussian-Mean problem \citep{2018A&A...616A...8A, fearnhead2024}. The CUSUM chart is known for its sensitivity to small shifts within a series of samples, thereby enabling the detection of jumps in star extinction features caused by the ISM. For the first time, our study applies CU-JADE method to identify extinction-jump points along the Distance-$A_\lambda$ ($D$-$A$) data sequence. 

In this section, we adopt the techniques in \citet{taylor2000change}, and improve them to suit our scenario. 
Considering a sequence of $N$ star samples exhibiting two distinct extinction-jump points in the $D$-$A$ space (Figure \ref{fig1}a), 
denoted as $A_{\lambda}$: $\{A_{1}, A_{2}, \cdots, A_{i}, \cdots, A_{N}\}$, along with the corresponding distance of each star $D$: $\{D_{1}, D_{2}, \cdots, D_{i}, \cdots, D_{N}\}$.  \\

Step 1. Get the first jump point. \\
Calculate the mean of all the $A_{\lambda}$, 
\begin{equation}
\bar{A} = \sum\limits^{N} A_{i}/N. 
\end{equation}
Begin by initializing the cumulative sum (CUSUM statistic; $S$) at zero, 
\begin{equation}
S_{0} = 0. 
\end{equation}
Sort the $A_{\lambda}$ in ascending order based on distance. Compute the CUSUM statistic $S$ for the remaining ordered values,
\begin{equation}
S_{i}=S_{i-1}+(\bar{A}-A_{i}), ~i>1. 
\end{equation}
For any given series of $A_{\lambda}$ values, it returns a sequence of cumulative sum values in this manner: $\{S_{1}, S_{2}, \cdots, S_{i}, \cdots, S_{N}\}$. 
In this case, for a group of $A_{\lambda}$ values in ascending order, the Distance-CUSUM $S$ ($D$-$S$) plot shows a trend where $S_{i}$ increases initially, and after reaching the peak it descends until it drops to zero (see red curve in Figure \ref{fig1}b). The prominent peak (denoted by a red star in the CUSUM curve) is preliminarily identified as the first extinction-jump point. 

The turning point is caused by the change in the difference between the $A_{\lambda}$ values and their mean $\bar{A}$. In the absence of extinction-jump points, or when the data sequence is randomly permuted, the CUSUM statistic $S$ would exhibit random fluctuations, characterized by the absence of statistically significant peaks or bulges (see colored curves in Figure \ref{fig1}c). This feature can help us determine whether the identified jump point is valid.  \\

Step 2. Validation of the identified jump point. \\
The confidence level ($CL$) regarding the existence of a jump point in the $D$-$A$ sequence can be assessed using $S_{\rm diff}$, which characterizes the prominence of the identified spike or turning point in the CUSUM analysis. This value is calculated as follows:

\begin{equation}
S_{\rm diff} = max\{S_{1}, S_{2}, \cdots, S_{i}, \cdots, S_{N}\} - min\{S_{1}, S_{2}, \cdots, S_{i}, \cdots, S_{N}\}. 
\end{equation}

In an ordered sequence of $A_{\lambda}$ values, which in our case increase monotonically with distance and exhibit some dispersion, the $S_{\rm diff}$ calculated above is set to $S^{0}_{\rm diff}$ (see Figure \ref{fig1}c). We shuffle the ordered sequence $N_{\rm shuf}$ times (typically $N_{\rm shuf}$=1000, empirically determined to ensure statistical robustness), thereby generating a new ensemble of $S_{\rm diff}$ values following the shuffling procedure. The $CL$ is then quantitatively defined as the percentile rank of $S^{0}_{\rm diff}$ relative to the null distribution $\{S_{\rm diff}^{1}, S_{\rm diff}^{2}, \cdots, S_{\rm diff}^{N_{\rm shuf}}\}$, as illustrated in panel (d) of Figure \ref{fig1}. 

To enhance the statistical robustness of the $CL$ and mitigate the risk of false positive detections, we introduce a scaling factor $C$ to modulate the statistic $S_{\rm diff}$: $\{C \cdot S_{\rm diff}^{1}, C \cdot S_{\rm diff}^{2}, \cdots, C \cdot S_{\rm diff}^{N_{\rm shuf}}\}$. The optimal $C$ value ($C=1-3$ is usually enough, $C=5$ in the Figure \ref{fig1}c and d) can be determined through validation testing, using both observational catalogs and mock data (see Section \ref{sec4}). 
Finally, the presence of a jump point is evaluated using a predefined $CL$ threshold, typically set at 0.95. \\


Step 3. If the jump point is valid, the sequence is split into two segments, and the iteration proceeds. \\
We divide the original sequence into two parts at the position of the first spike. The same procedure is then applied recursively to each subset until the $CL$ of a subset falls below the predefined threshold. \\

Step 4. The last jump point: a single-extinction-jump scenario. \\
During the iteration, from panel (b) to (f) in Figure \ref{fig1}, we find that the CUSUM statistic $S^{0}_{\rm diff}$ of an ordered sequence drops significantly. We attribute the progressive decline in $S^{0}_{\rm diff}$ during Step 3 to the variation in the jump value $\Delta A_{\lambda}$, the number of stellar sample, and the caveats from early-detected jump points. The last factor indicates that the first detected jump points are not necessarily the most significant ones and may carry elevated false-positive risks. Focusing on the final set of detected jumps (our termination condition) might simplify our understanding for interpreting multijump scenarios. In the next subsection, we quantitatively analyze factors influencing the $CL$ of jump point detection. For extended discussion and complications, see the Appendix \ref{subsec:a1}. \\

Steps 5--6. Return all jump points and check against input/external constraints. \\
These two steps primarily serve to validate the method's accuracy and optimize parameter selection. A more comprehensive analysis is presented in Section \ref{sec4}.


\subsection{Detection of Weak Jumps in D-A Plot} \label{subsec:33}
Considering the end of the iterative procedure in Steps 3 and 4 (as an example of Figure \ref{fig1}f), for a set of $D$-$A$ data with a single jump point, we can obtain $S_{\rm diff}^{0}$ according to the definition
\begin{equation}
S_{\rm diff}^{0} \propto N ~\Delta A_{\lambda}. 
\end{equation}
Obviously, the prominence of a jump point is proportional to the number of samples $N$ and the magnitude of the jump $\Delta A_{\lambda}$. For two sample groups with a known distribution, the jump values $\Delta A_{\lambda}$ are considered fixed parameters. Consequently, the number of samples is the primary factor affecting the prominence of the jump point . 

Assessing $CL$ requires an in-depth understanding of how many factors affect the prominence $S_{\rm diff}$ of the shuffled samples. We assume that $S_{\rm diff}$ is dependent on the number of stars $N$, the magnitude of the jump value $\Delta A_{\lambda}$, and the dispersion of the $A_{\lambda}$ values, denoted as $\sigma$. In this work, the sample size $N$ reflects the number density $n$ of stars when the distance range is specified. Through analytical derivation and numerical computation (see details in Appendix \ref{subsec:a2}), we have 

\begin{equation}
S_{\rm diff} \propto \sqrt{n}~ \xi\left(\Delta A_{\lambda}, \sigma\right). 
\end{equation}

$\xi\left(\Delta A_{\lambda}, \sigma\right)$ exhibits a complex analytic form. However, $\xi\left(\Delta A_{\lambda}, \sigma\right) \propto \sigma$ when $\sigma \gtrsim \Delta A_{\lambda}$. 
We focus on analyzing two primary cases. When $\Delta A_{\lambda} \gg \sigma$, we define these features as significant extinction-jump points. The algorithm facilitates efficient traversal of all jump points on the $D$-$A$ diagram (see examples in Figure \ref{fig2}).  
For the other case, due to intrinsic dispersion and uncertainty, the extinction of stars in Gaia data leads to a magnitude uncertainty of $\sigma$ that is comparable to most of the $\Delta A_{\lambda}$ values we can detect as jump points. 
In observations, these jump points are collectively referred to as ``weak jumps'', and statistically the number of these weak jumps accounts for the majority. Additionally, the $D$-$A$ diagram includes gentle slopes ($k$) that may be caused by diffuse and extended gas along the LoS, which exhibit more linear properties in $D$-$A$ space and present challenges for the CU-JADE method (see Appendix \ref{subsec:a3}). 
In the following subsections, we will address the influence of these factors on the results and optimize the parameter settings to identify these extinction features with weak jumps. 

\subsubsection{Comparable $\Delta A_{\lambda}$ and $\sigma$ for Weak Jumps} \label{subsec:331}
By Equations (6), (A7) and Figure \ref{fig:appa}, when $\Delta A_{\lambda}$ and $\sigma$ are the same magnitude, $\sigma$ has a major influence on the $S_{\rm diff}$. 
Approximately, we have $S_{\rm diff}~\propto~\sqrt{n}~\sigma,~\sigma \gtrsim \Delta A_{\lambda}$. 
The influence of $\Delta A_{\lambda}$ on $S_{\rm diff}$ is negligible compared to $S_{\rm diff}^{0}$. Combining Equation (5), we have 
\begin{equation}
CL \propto \sqrt{n}~ \Delta A_{\lambda}~ \sigma^{-1}. 
\end{equation}

The $CL$ for an extinction-jump is proportional to the jump value and the square root of number density, but inversely proportional to $A_{\lambda}$ dispersion. Therefore, a larger sample size tends to make it easier to detect a jump point, as well as the larger $\Delta A_{\lambda}$.
A large dispersion would smear out the jump when it is comparable to the $\Delta A_{\lambda}$ values. Since $\Delta A_{\lambda}$ and $\sigma$ depend upon the LoS extinctions and the uncertainties of data, the value of $CL$ now is only influenced by the number density of the samples in the calculation. 

\subsubsection{Two Modes for Stellar Sample Density Variation} \label{subsec:332}

When selecting a group of stars along the LoS, the $n$  is proportional to angular area. This raises a problem: a higher number density might introduce more jump points, as the $CL$ also increases. At the same time, larger angular area might include other components in different distances along the LoS. We propose two implementation modes for different scenarios: the first (Mode I) fixes the $CL$ threshold, meaning that the chosen number of star samples directly affects detection. Mode I is good at detecting extinction jumps in an unknown area and is not sensitive to low number density. We use this mode to search for the extinctions associated with HMSFRs and traverse all-sky extinction jumps in Sections \ref{validatemaser} and \ref{dustmap}, respectively. 

The second mode (Mode II) adjusts the $CL$ threshold with the number density, scaling with $\sqrt{n}$, to mitigate the influence of star density. In Mode II, we refine our method to apply varying $CL$ thresholds in each split iteration, enabling the jump point detection in intervals with sparser star distribution. 
In cases with prior knowledge of the target area, we recommend the Mode II. It is important to control bias stemming from sample density. The number density $n$ is calculated based on the star samples number per unit length in a specific region and distance interval, $n = \frac{N}{d_{max}-d_{min}}$, where $N$ is the number of stars, $d_{max}$ and $d_{min}$ are 
the maximum and minimum distances of selected stars in each interval, respectively. 

As discussed in Section \ref{subsec:31}, to enhance the $CL$ while mitigating false detections, we multiply $S_{\rm diff}$ by a factor $C$ ($C\sqrt{\beta n}$ for Mode II, $\beta$ is a constant for fine-tuning).  
$C$ is usually set to be $1-3$, considering $\Delta A_{\lambda}$ ranges from $\left[0.5, 5\right]$ mag and $n$ ranges from $\left[0.1, 10\right]~{\rm pc}^{-1}$. In the cases of large dispersion $\sigma$, or too small $\Delta A_{\lambda}$ due to weak extinctions, we have to do coarse tuning to the criteria by reducing the value of $C$, as $CL$ decreases rapidly during the iterations. We find that $C=1$ is more effective for weak jumps (see details in Section \ref{mock}). Fine-tuning of the threshold can also be achieved by altering the percentile value of $S_{\rm diff}^{0}$ in shuffled $S_{\rm diff}$ (generally taken as 0.95). 
Mode II has to be complemented with both a lower and an upper limit. That is, samples with few stars cannot reveal extinction structures, while larger samples may detect weak jumps and additional extinctions from other components along the LoS. We establish an interval for $n$ to perform density adjustment, and samples outside of this interval will be fixed at a predefined lower or upper bound. 
Considering that a typical MC with an angular size $\rm \gtrsim 10-50~ arcmin^{2}$ \citep{2021ApJS..257...51Y, 2024AJ....167..220Z} can be measured within 3 kpc near the Galactic plane, we recommend a range of the number density of stars as $n \in \left(0.2, 20\right) {\rm pc}^{-1}$ and that the iteration stops when the number of stars in a subset is less than 20. 

\subsection{The Uncertainties of Distance by Bootstrap} \label{boot}

Since the uncertainties of the distances of the jump points cannot be directly derived from the data uncertainties, the bootstrap method offers a way to reflect systematic measurement errors in our results.  A bootstrap is described by $X_{i} = \mu_{i}+\varepsilon$, where $X_{i}$ is the new dataset used for calculations, $\mu_{i}$ is the original value corresponding to distance or extinction in this study, and $\varepsilon$ represents the small perturbations.

For data with small errors, we simplify this as a Gaussian distribution ($\mu_{i}$ for the median value and $\varepsilon$ for the standard deviation). We generate 1000 new datasets based on the error associated with each data point, combining method uncertainties and systematic errors from the observation data. For instance, we rank the results and use the 16th and 84th percentiles of the ordered positions to define the lower and upper limits of the 0.68 confidence interval. We adopt the median value of the jump point as the final estimated distance derived from our method. 

Here, we can primarily define an extinction-jump as an increase in extinction values over a short distance interval. For instance, under typical distance uncertainties of 5\%--10\% ($\delta D \sim 50$ pc), $\Delta A_{V}$ reaches $\sim 0.15$ mag (see Section \ref{dustmap}). The ratio between $\Delta A_{V}$ and $\delta D$ reflects a density of extinction, equivalent to $k \gtrsim 0.003 ~{\rm mag \cdot pc^{-1}}$ (see Appendix \ref{subsec:a3}). 
Obviously, the definition of jumps is mainly limited by the uncertainties of star samples. Fortunately, in real data, the increase in extinction \citep[$\sim 0.75~{\rm mag \cdot kpc^{-1}}$;][]{1982A&A...109..213L, 2005A&A...436..895G} brought by the slope $k$ is generally less than the extinction feature caused by jumps. 

\begin{figure*}[htbp]
\centering
\includegraphics[width=18cm,angle=0]{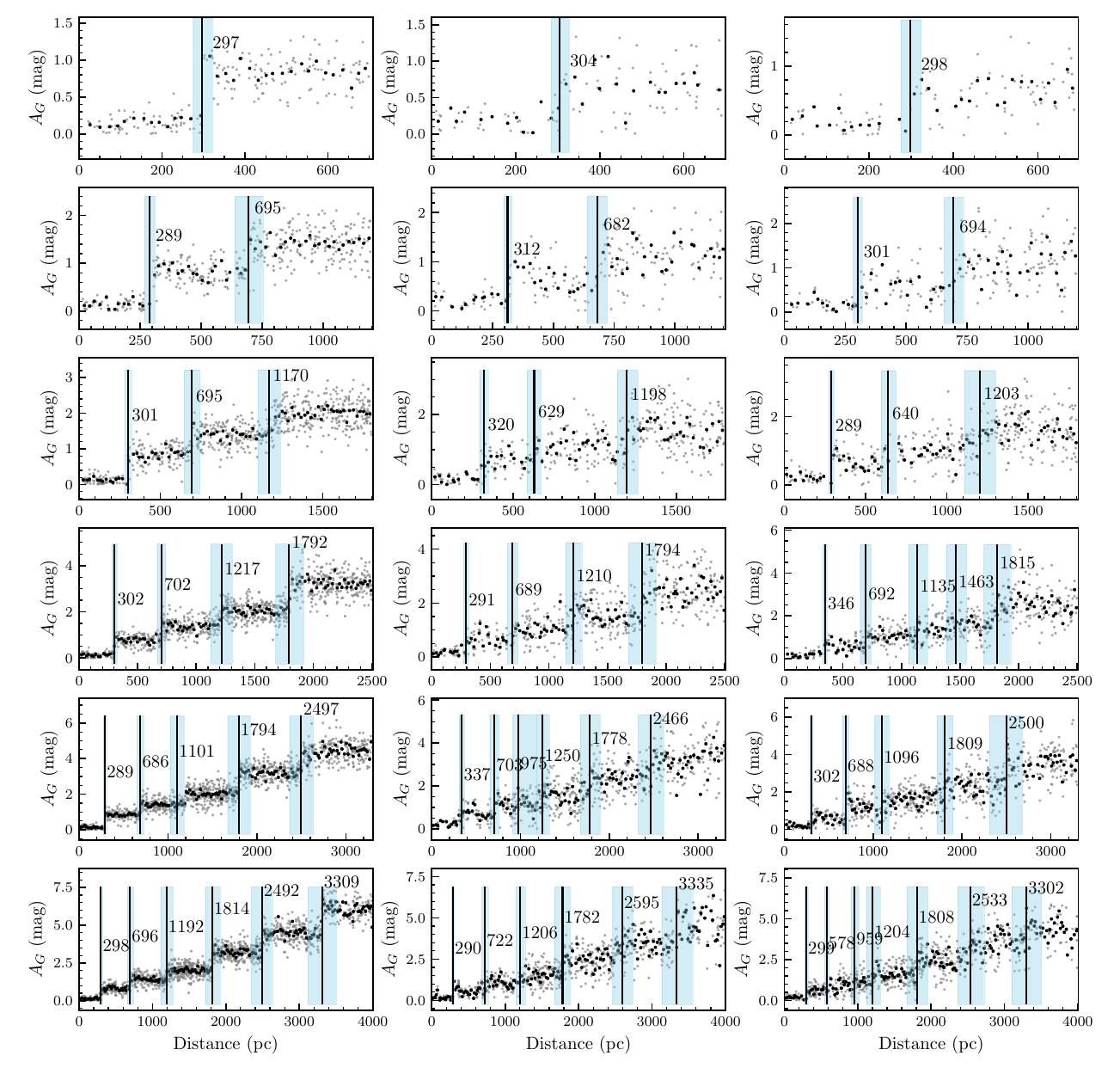}
\caption{An illustration of our multi-extinction-jump model (here in Mode I, $C$=3 for the left column, $C$=1 for the middle and right columns) applied to the mock data. Different rows set the jump points number from 1 to 6, with distances at $\left[300, 700, 1200, 1800, 2500, 3300\right] {\rm pc}$, respectively. The three columns mirror different LoS extinction properties (see the text). Gray dots are the produced mock data and black dots are binned values every 20 pc.  \label{fig2}}
\end{figure*}

\section{Validation} \label{sec4}
\subsection{Validation Using Mock Data} \label{mock}
To verify the ability of our method to detect weak jumps, we generate mock data to simulate the distribution of stars in the $D$-$A$ diagram. 
Firstly, we categorize the synthetic extinctions into different distance intervals with distinct mean and dispersion. Both the distribution of distance errors and $A_{G}$ errors follow a truncated Gaussian model. Referring to stars in Cygnus (from Gaia DR3 in Section \ref{gaiadr3}), we found that the distance error is linearly correlated with distance, with a coefficient of approximately 0.1 and a dispersion of about 0.06. 
As for $A_{G}$ errors, their average and dispersion show no apparent correlation with $A_{G}$ itself. We have therefore set them to 0.04 mag for both the average and the dispersion based on the data toward the Cygnus. 

In Figure \ref{fig2}, the number of jump points increases from one to six in different rows from top to bottom. For the columns, the left column presents a simple multijump pattern, resembling a stairs with different star number densities
$n = \left[0.2, 0.3, 0.4, 0.5, 0.4, 0.3, 0.2\right] {\rm pc}^{-1}$. The middle column gives smaller $\Delta A_{G}$ and 
larger $\sigma$ compared to left panels with lower number density $n = \left[0.1, 0.15, 0.2, 0.25, 0.2, 0.15, 0.1\right] {\rm pc}^{-1}$. 
The panels in right column have an added a linear gradient with a slope $k\sim 3$ $\times 10^{-5}$ $\rm mag \cdot pc^{-1}$ based on the middle column. The distances corresponding to each jump point are set to $\left[300, 700, 1200, 1800, 2500, 3300\right] {\rm pc}$. Obviously, our model demonstrates good performance on these mock data (see the identified extinction jumps indicated by vertical lines in Figure \ref{fig2}). 

Through statistical analysis of large number of mock data, we are able to quantitatively evaluate the optimized parameters of our method for various application scenarios. We set up three types of mock data (1, 2, and 3 jump points), with 300,000 groups of $D$-$A$ data sequences for each type. In addition to the uncertainties of distance and extinction setting before, we randomly combine the following settings for $D$, $\Delta A_{G}$, $\sigma$, $n$, and $k$ to obtain $D$-$A$ data sequences. These $D$-$A$ data sequences represent that the physical quantities are completely randomly generated and combined under a certain statistical distribution. For weak jumps with a linear extinction gradient along the LoS, we set up (1) stars' distances in [10, 3000] pc; (2) jump points' distances uniformly distributed in [60, 2950] pc; (3) $\Delta A_{G} \sim \xi(0.2)+0.15$ mag, where $\xi$ represent an exponential distribution; (4) $\sigma \sim \xi(0.2)+0.2$ mag; (5) $P(n|0.2,0.05)$ is Gaussian for $n$ in [0.1, 0.4] $\rm pc^{-1}$; (6) $P(k|3,1)$ is Gaussian for $k$ in [1, 6]$\times 10^{-5}$ $\rm mag \cdot pc^{-1}$. 

By adjusting the parameter $C$, we analyze the three types of mock data to understand their behavior under various conditions. The detection of jump point  in the mock data is categorized into the following three situations: detecting the preset jump points (True Positive; $TP$); failing to detect the preset jump points (False Negative; $FN$); and detecting non-preset change points (False Positive; $FP$). We determine whether a jump point has been successfully detected by seeing if the difference from the preset value is within the $1\sigma$ confidence interval. The reliability ($precision$) of detection is the proportion of $TP$ among all detected change points $presicion=\frac{TP}{TP+FP}$, and the completeness ($recall$) is the proportion of $TP$ among all preset jump points $recall=\frac{TP}{TP+FN}$. The $F1$ score, which serves as a metric for the model's accuracy in classification, can be derived from $precision$ and $recall$ $F_{1}=2\times\frac{precision \times recall}{precision + recall}$. 

Table \ref{tab1} lists the mean of all the statistical results, the high $precision$, $recall$, and $F1$ score presents good performance for detecting weak jump points with slopes for $C=1$. Raising $C$ will improve the $precision$, but the $recall$ decreases significantly. In addition, the mean $F1$ score of the first detected jump point shows no significant difference compared to the scores of all detections. This indicates that early-detected jump points also exhibit robustness, while the number of jump points has negligible impact on the jump point detection, even though the downward trend of $F1$ score with an increasing number of jump points reveals some issues (see Appendix \ref{subsec:a1}). The hypothesis testing with mock data indicates that controlling a single parameter ($C$) helps this method to detect fewer erroneous change points. 

\begin{table} [!htbp]
\tablenum{1}
\caption{The statistic of CU-JADE method for weak jumps with slopes along the LoS.}
\begin{tabular}{cccccccccc}
\toprule
$C$ & $\rm Type^{a}$ & $TP$ & $FP$ & $FN$ & $Precision$ & $Recall$ & $F1$ score & $\rm Gap^{b}$ & $F1$ $\rm score_{I}^{c}$ \\
\midrule
\multirow{3}{*}{1} & 1 & 0.8418 & 0.3694 & 0.1582 & 0.6950 & 0.8418 & 0.7614 & 0.2384 & 0.7314 \\
& 2 & 1.5902 & 0.4286 & 0.4098 & 0.7877 & 0.7951 & 0.7914 & -0.0245 & 0.6129 \\
& 3 & 2.3144 & 0.4719 & 0.6856 & 0.8306 & 0.7715 & 0.8000 & -0.4094 & 0.5313 \\
\midrule
\multirow{3}{*}{2} & 1 & 0.6470 & 0.0420 & 0.3530 & 0.9390 & 0.6470 & 0.7661 & -0.3081 & 0.7525 \\
& 2 & 1.1738 & 0.0874 & 0.8262 & 0.9307 & 0.5869 & 0.7198 & -0.8448 & 0.6190 \\
& 3 & 1.6921 & 0.1076 & 1.3079 & 0.9402 & 0.5640 & 0.7050 & -1.4825 & 0.5410 \\
\midrule
\multirow{3}{*}{3} & 1 & 0.4070 & 0.0033 & 0.5930 & 0.9919 & 0.4070 & 0.5772 & -0.5896 & 0.5769 \\
& 2 & 0.7424 & 0.0240 & 1.2576 & 0.9687 & 0.3712 & 0.5367 & -1.3325 & 0.5216 \\
& 3 & 1.0698 & 0.0390 & 1.9302 & 0.9648 & 0.3566 & 0.5207 & -2.1363 & 0.4848 \\
\bottomrule
\end{tabular}
\tablenotetext{Note. } 
{\\ $^{a}$ Three types of mock data, with 1, 2, and 3 preset jump points, respectively. 
\\ $^{b}$ Gap refers to the average number of detected jump points minus the preset number.  
\\ $^{c}$ The $F1$ score of the first detected jump point.
}
\label{tab1}
\end{table}

\subsection{Validation Using Distances of Masers} \label{validatemaser}
The dense molecular clumps associated with masers can cause significant stellar extinction. 
When applying the CUSUM-based model to stars along the LoS toward masers, we expect to detect jumps in $D$-$A$ that are likely caused by the associated MCs. 
We use maser samples from \citet{2019ApJ...885..131R} and \citet{2023ApJ...953...21H} to correspond the distance of HMSFRs. 
In total, over 200 of masers have had their parallaxes and proper motions measured. Considering the precision of Gaia's geometric measurements of stars, we selected 75 maser samples with parallaxes greater than 0.333 mas for validation. 

We selected all stars with a relative parallax error smaller than 0.2, centered on the locus of each maser within a projection radius of  $0^{\circ}.3$. The maximum distance cutoff was set at 5 kpc. We applied the Mode I method, which involves setting a universal constant $C=3$. Here, the scaling factor $C=3$ is quite strict for identifying the jumps caused by the majority of these HMSFRs traced by maser measurements, with $n$ in the range of $\left(0.2, 0.6\right) {\rm pc}^{-1}$. We manually match the distances derived from maser parallaxes with the locations where $A_{G}$ exhibited jumps. One may wonder whether the matching of jump points to maser distances is due to human factors.
In the majority of our selected samples, there are only two or three jump points are observed within 5 kpc. Therefore, it is reasonable to consider that the association between the identified jump points and masers with the corresponding HMSFRs is not coincidental (see examples in Figure \ref{figc1}). 
We conclude that our method can independently capture the dense gas structures with $\Delta A_{G} \gtrsim 0.5$ mag based on the well crossmatch between the extinction distances and masers parallax distances. 

\begin{figure}[htbp]
\centering
\includegraphics[width=16cm,angle=0]{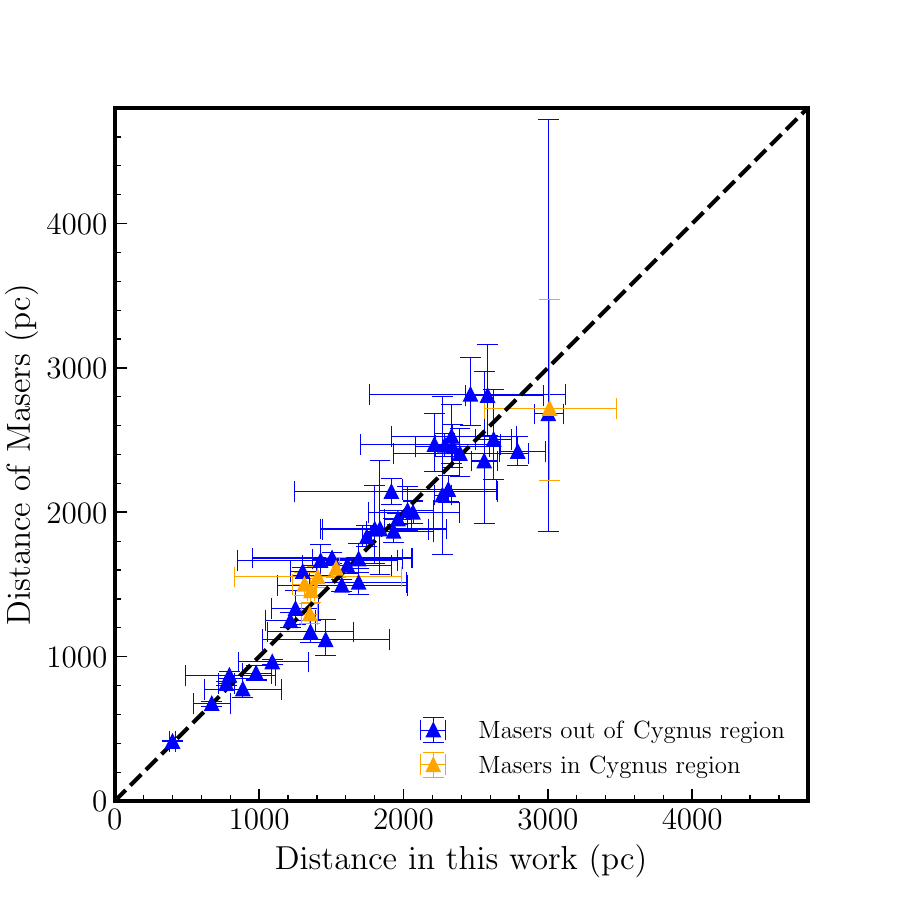}
\caption{The distance correspondence between parallaxes from masers and those derived from our multi-extinction-jump model. Orange triangles highlight masers in the Cygnus region, while blue triangles represent those outside the Cygnus region. \label{fig3}}
\end{figure}

Finally, 46/75 (42/58) of masers within 3 kpc (2.5 kpc) are found to correspond with the jump points identified by our Mode I. 
A good linear relationship between all the detected distances and those derived from masers are well presented in Figure \ref{fig3} without systematic deviation. Most of the differences between two methods fall within the statistical tolerance of a $1\sigma$ error. 
Among them, for some samples with small number density, we lowered the scaling factor appropriately and successfully recovered another six samples (five cases with $C=2$ and one with $C=1$). 
The consistency in the figure indicates that our method can well trace the distance of the dense regions of HMSFRs. 

Our method fails to detect approximately 28\% of the sample within 2.5 kpc and 76\%  within 2.5--3 kpc range. This is attributed to a significant decrease in the completeness of stars with precise parallax and $A_{G}$ measurements, particularly due to 
the ``extinction wall'' phenomenon \citep{1993BaltA...2..171S}, where foreground dust obscures distant stars, especially those 
behind the Aquila Rift and Cygnus Rift. Notwithstanding these limits, in the Cygnus region, our method successfully identified multiple extinction jumps associated with MCs linked to masers in star-forming regions beyond the Cygnus Rift (highlighted as orange points in Figure \ref{fig3}). This method proves more effective than the Bayesian single-extinction-jump method \citep{2024AJ....167..220Z} in detecting extinction components at greater distances with fewer background stars. Furthermore, as distance increases, especially beyond 2.5 kpc, the inability to detect the more distant extinction-jump points is not due to
mismatches with maser distance measurements but rather because of insufficient stellar sample density in these regions. 


\section{Implementation of the method} \label{sec5}

\subsection{All Sky Dust Map} \label{dustmap}
\begin{figure*}[htbp]
\centering
\includegraphics[width=18cm,angle=0]{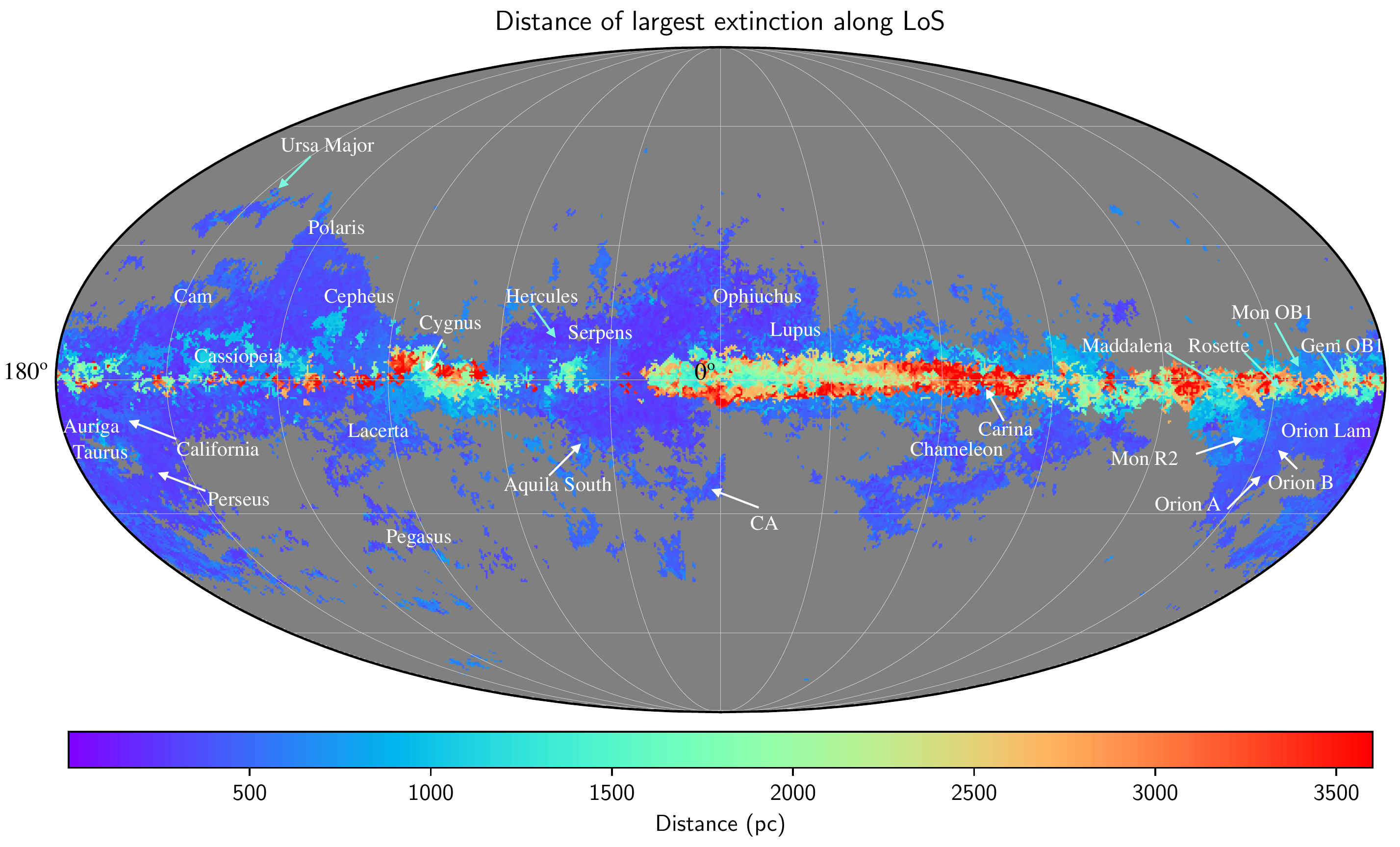}
\caption{The sky map displays the distances corresponding to the largest $\Delta A_{V}$ component along the LoS in a Mollweide view. Some well-studied regions are indicated on the map. It should be noted that the values are likely to represent the MCs' distances, but they could also correspond to other components superimposed on them. For the slices (every 200 pc) and integrated extinctions map for 0--4 kpc, please check the \href{https://www.scidb.cn/file?fid=940f358ac99ec82696f71448ded54e1f&mode=front}{10.57760/sciencedb.24650}. 
\label{fig:edgeonmap}}
\end{figure*}

This method does not require prior assumptions of the number of jump points. Indeed, due to the complex distribution of gas on the Galactic plane, it is difficult to preset how many extinction components there are along a certain LoS. 
For instance, the $D$-$A$ curve is not merely a simple step function, but involves combinations of different scenarios (see details in Appendix \ref{app:appb}). Our method can traverse the jump points along the LoS in a certain region. This can be applied to detect components with significant extinction across the full sky, effectively generating a 3D extinction map for whole sky (see Figures \ref{fig:edgeonmap} and \ref{fig:polarmap} for the sky map and top-down views). 

To map all LoS characteristics across the celestial sphere, we apply the cylindrical equal area rectangular projection, which projects the sphere onto a rectangular plane. We set the sampling size (represent spatial resolution) $L_{0}=1^{\circ}$ and sampling interval (grid size) $L=30'$ for a guide map. Setting aside the notable variations in star density across high Galactic latitudes and the Galactic plane, we generate a uniformly sampled planar diagram with a $30'$ grid spacing and $30'$ overlap to enhance the spatial correlation. This standardized format facilitates easy comparison with other maps. However, unlike traditional extinction maps, the distribution of extinction jumps in our map is discrete in the dimension of distance. For the sky map, we perform integration across various distance slices, while for the face-on map, we employ a 2D histogram with extinction magnitude as the weighting factor. 

\subsubsection{A Dust Map based on SHEDR3} \label{shemap}

We utilized the dataset from SHEDR3 (by Mode I, $C=2$), providing measurements of extinction $A_{V}$ and distance, with the maximum star distance capped at 5 kpc. The integration of this star table with infrared data aids in detecting more distant components with deeper extinction. The magnitudes of all identified jump points adhere to an exponential distribution. To improve detection accuracy and focus on more significant jumps, 
we retain jumps with $\Delta A_{V}$ exceeding 0.15 mag \footnote{
A sample of this catalog with the description of the columns can be found in Table \ref{tab2}. The full machine-readable table of extinction components is presented in  \href{https://www.scidb.cn/file?fid=082d6e61c31a6d867b9fe8aedfbdfc50&mode=front}{10.57760/sciencedb.24650}, from which the 3D extinction maps and distances of MCs can be derived.}.
This threshold is set by considering three key reasons: (1) 0.15 mag represents a typical precision value for extinction in the $V$ band from SHEDR3 at $G=17$ mag \citep{2022A&A...658A..91A}; (2) the simulations in section \ref{mock} demonstrate the algorithm's effectiveness in detecting jump points above this threshold (assuming $\sigma \sim 0.2-0.5$ mag); (3) The most important is that, at this threshold, molecular hydrogen might be experiencing a phase transition based on the conversion between the extinction magnitude and the total column density of all gas. 
Here, we apply the empirical relationship between HI and the reddening, given by $N_{\rm HI}/E(B-V)=8.8\times10^{21}{\rm cm}^{-2}{\rm mag}^{-1}$ \citep{2017ApJ...846...38L}, with an extinction coefficient $R_{V}=3.1$ \citep{1989ApJ...345..245C, 2011ApJ...737..103S}. This implies that the extinction contributed by HI is $A_{V}({\rm HI})=0.35\times10^{-21}N_{\rm HI}$. And the total extinction $A_{V}$ reflects a combination of atomic gas and molecular hydrogen, which is $N_{\rm H}=N({\rm HI})+2N({\rm H}_{2})\approx1.9\times10^{21}{\rm cm}^{-2}A_{V}$ \citep{1978ApJ...224..132B}. Based on empirical relationships, the column density range where molecular hydrogen appears is $log~N_{\rm H}({\rm cm}^{-2})\approx 20.1-20.8$ \citep{2006ApJ...636..891G, 2020A&A...643A..36B}, corresponding to an extinction value of $A_{V}({\rm mag})\approx 0.06-0.3$, with 0.15 mag being a mid-range value. Concurrently, \citet{2001ApJ...547..792D} suggests an extinction threshold for CO of $\gtrsim 0.4$ mag, indicating that some CO-dark molecular gas might appear below $\sim 0.4$ mag \citep{2020A&A...639A..26K, 2024A&A...682A.161S}. We expect that a higher sensitivity survey may detect weak CO emission in some CO-dark regions. 

\begin{table}[htbp]
\tablenum{2}
\caption{A sample of the all-sky extinction catalog derived from SHEDR3 (completely given in electronic form).}
\label{tab2}
\resizebox{0.86\textwidth}{!}{
\begin{tabular}{cccccccccccc}
\toprule
No. & 
$l$ & 
$b$ & 
$N_{\mathrm{comp}}$ & 
$D_{50}$ & 
$D_{16}$ & 
$D_{84}$ & 
$\mu^{\mathrm{fore}}$ & 
$\mu^{\mathrm{back}}$ & 
$\sigma^{\mathrm{fore}}$ & 
$\sigma^{\mathrm{back}}$ & 
$\Delta A_{V}$ \\
 & (deg) & (deg) &  & (kpc) & (kpc) & (kpc) & (mag) & (mag) & (mag) & (mag) & (mag) \\
\midrule
1  & 0.0 & -25.5 & 3 & 0.441 & 0.422 & 0.501 & 0.279 & 0.446 & 0.168 & 0.185 & 0.168 \\
2  & 0.0 & -25.0 & 3 & 0.380 & 0.368 & 0.389 & 0.255 & 0.421 & 0.157 & 0.161 & 0.165 \\
3  & 0.0 & -24.5 & 3 & 0.379 & 0.368 & 0.389 & 0.250 & 0.405 & 0.162 & 0.162 & 0.155 \\
$\vdots$ & $\vdots$ & $\vdots$ & $\vdots$ & $\vdots$ & $\vdots$ & $\vdots$ & $\vdots$ & $\vdots$ & $\vdots$ & $\vdots$ & $\vdots$ \\
152563 & 359.5 & 30.5 & 2 & 0.332 & 0.319 & 0.391 & 0.553 & 0.796 & 0.266 & 0.191 & 0.243 \\
152564 & 359.5 & 42.5 & 1 & 0.539 & 0.455 & 0.745 & 0.474 & 0.630 & 0.195 & 0.163 & 0.157 \\
\bottomrule
\end{tabular}}
\tablenotetext{\rm The ~description ~of ~the ~columns. } 
{\small
\\ $l$: The Galactic longitude coordinate of extinction components across the sky. 
\\ $b$: The Galactic latitude coordinate of extinction components across the sky.  
\\ $N_{\rm comp}$: The number of LoS extinctions identified by the CU-JADE method, including those with $\Delta A_{V}<0.15$ mag. 
\\ $D_{50}$: Median distance obtained from bootstrap sampling, quoted as the determined distance. 
\\ $D_{16}$: The 16-th distance obtained from bootstrap sampling, quoted as the lower limit of distance in $1 \sigma$ confidence interval. 
\\ $D_{84}$: The 84-th distance obtained from bootstrap sampling, quoted as the upper limit of distance in $1 \sigma$ confidence interval. 
\\ $\mu^{\mathrm{fore}}$: The mean extinction in the prejump distance bin. 
\\ $\mu^{\mathrm{back}}$: The mean extinction in the postjump distance bin. 
\\ $\sigma^{\mathrm{fore}}$: The standard deviation of extinction in the prejump distance bin. 
\\ $\sigma^{\mathrm{back}}$: The standard deviation of extinction in the postjump distance bin. 
\\ $\Delta A_{V}$: The magnitude of extinction jumps identified by the CU-JADE method. 
}
\end{table}

In Figure \ref{fig:edgeonmap}, the Mollweide projection plot of the sky map illustrates the distances associated with the most significant extinction jumps across all LoS. The diagram shows some notable nearby regions that likely match those identified MCs by the single-extinction-jump model \citep{2019ApJ...885...19Y, 2019ApJ...879..125Z}. 
The advantages of our model are that (1) the model with multi-extinction-jump 
can well trace all extinction structures with considerable $\Delta A_{\lambda}$
and (2) all the detected dust components are unaffected by different modeling process, i.e., the ``ramps'' in the $D$-$A$ map \citep{2019ApJ...879..125Z} and the overlapping between multiple gas layers in distance measurements \citep[i.e., multiple jump features along LoS, see][]{2024AJ....167..220Z}. 
For the majority of the LoS, particularly at high Galactic latitudes, MCs are usually isolated and correspond to the strongest extinction components, facilitating the identification of their associated extinction structures and distances \citep{2024AJ....168..203S}. 
In contrast, the Galactic plane often features multiple molecular gas layers associated with various extinction components, complicating the precise determination of their distances. Given the correlation between MC column density and extinction, we prioritize the analysis of more significant extinction components (see Sections \ref{cep} and \ref{cyg}). 

\begin{figure*}[htbp]
\centering
\includegraphics[width=18cm,angle=0]{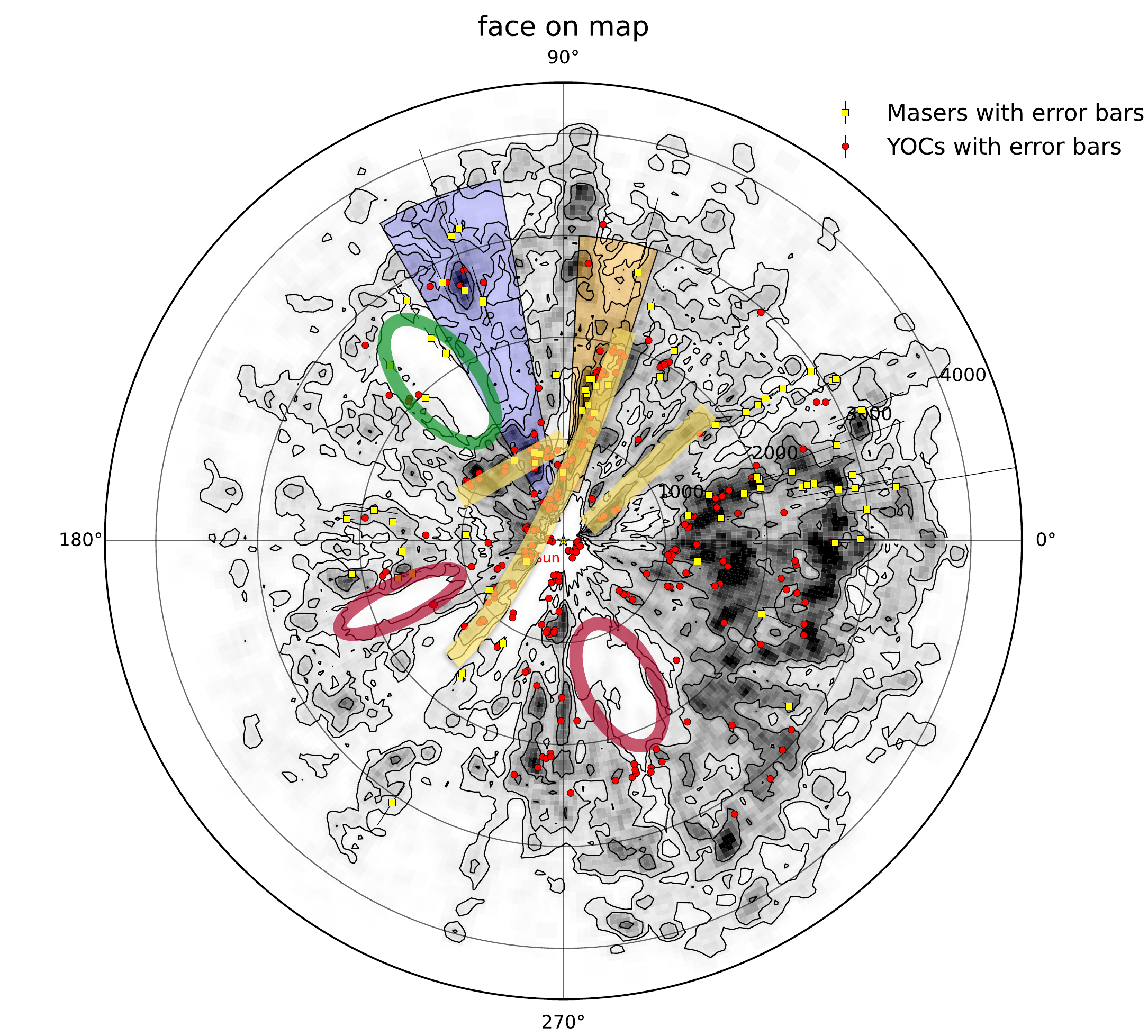}
\caption{A 2D histogram illustrating $\Delta A_{V}$ weighted extinction jumps is presented in polar coordinates. The map has been smoothed using a uniform filter in Scipy with a kernel size of 3. The blue sector represents Cepheus, and the orange sector is Cygnus. Yellow squares with error bars represent masers, the data for which is sourced from \citet{2019ApJ...885..131R}. Red circles with error bars denote YOCs from \citet{2023A&A...673A.114H}. The light yellow belts delineate the kpc-scale linear structures \citep[e.g., the Split, the Radcliffe Wave, and the Cepheus spur in][]{2019A&A...625A.135L, 2020Natur.578..237A, 2021MNRAS.504.2968P}, while the red (the Mon-Gem superbubble and Carina superbubble) and green \citep{2022A&A...664A.174V} ellipses show large-scale cavities where dust is scarce. 
For more top-down views along different latitude slices, please refer to the \href{https://www.scidb.cn/file?fid=71a1f9da0dda2650dd02e6ffcb68b1f7&mode=front}{10.57760/sciencedb.24650}. 
\label{fig:polarmap}} 
\end{figure*}

In the polar coordinates (top-down veiw in Figure \ref{fig:polarmap}),  the line toward the Galactic Center is set at $0^{\circ}$, with the radius $r$ representing the distance from the Sun. We use $\Delta A_{V}\cdot r$ as a weighting factor (in Section \ref{zxymap}, $\Delta A_{V}\cdot r^{2}$ serves as the weight in the Cartesian coordinate system), making the linear colormap directly proportional to the surface density. The units for this scale are arbitrary and not specified here. 
In Figure \ref{fig:polarmap}, we have marked masers and YOCs with parallaxes $\geq0.3$ mas, demonstrating a good correspondence between the identified extinction structures and these dense gas regions. 
The blue and orange sectors represent two special cases located at high Galactic latitude region (Section \ref{cep} for Cepheus) and Galactic plane region (Section \ref{cyg} for Cygnus), respectively. 

Recent studies utilizing Gaia parallaxes have shed new light on the structure of the solar neighborhood. For example, some nearest star-forming regions are found to be located near the surface of the Local bubble \citep[e.g., Taurus, Ophiuchus, Lupus, Chamaeleon, and Corona Australis; see details in][]{2022Natur.601..334Z}. 
Our results also show prominent extinction features toward these regions within 300 pc (see thick blue regions in Figure \ref{fig:edgeonmap}). 
Some intriguing extinction structures are also revealed by our map (see light yellow bands for kpc-scale linear structures in Figure \ref{fig:polarmap}), e.g., the spur-like structure between the Sagittarius and Local arms \citep[``Split'' in][]{2019A&A...625A.135L},  
the Radcliffe Wave \citep{2020Natur.578..237A}, the Cepheus spur \citep{2021MNRAS.504.2968P}, etc. Besides, the large cavity in $120^{\circ} \lesssim l \lesssim 135^{\circ}$ \citep[green ellipse in Figure \ref{fig:polarmap},][]{2022A&A...664A.174V} is also clear in our map. We also find more oval cavity structures that dust is scarce in $200^{\circ} \lesssim l \lesssim 210^{\circ}$, $280^{\circ} \lesssim l \lesssim 300^{\circ}$ (see red ellipses in Figure \ref{fig:polarmap}). We call them the Mon-Gem superbubble and the Carina superbubble, respectively. While conventional extinction mapping obscures these features due to severe LoS extinction and artificial structures, our analysis more clearly reveals their kpc-scale morphology. These giant bubbles are reminiscent of the vast cavities observed across NGC 628 \citep[the Phantom Voids;][]{2023ApJ...944L..24W, 2023ApJ...944L..22B}, likely reflecting the global impact of intense past star formation activity on the gas distribution and dynamics of the Galactic disk. 
Future multiwavelength (e.g., X-ray and infrared) observations will be essential for further verification and to confirm their formation mechanisms.

\begin{figure*}[htbp]
\centering
\includegraphics[width=18cm,angle=0]{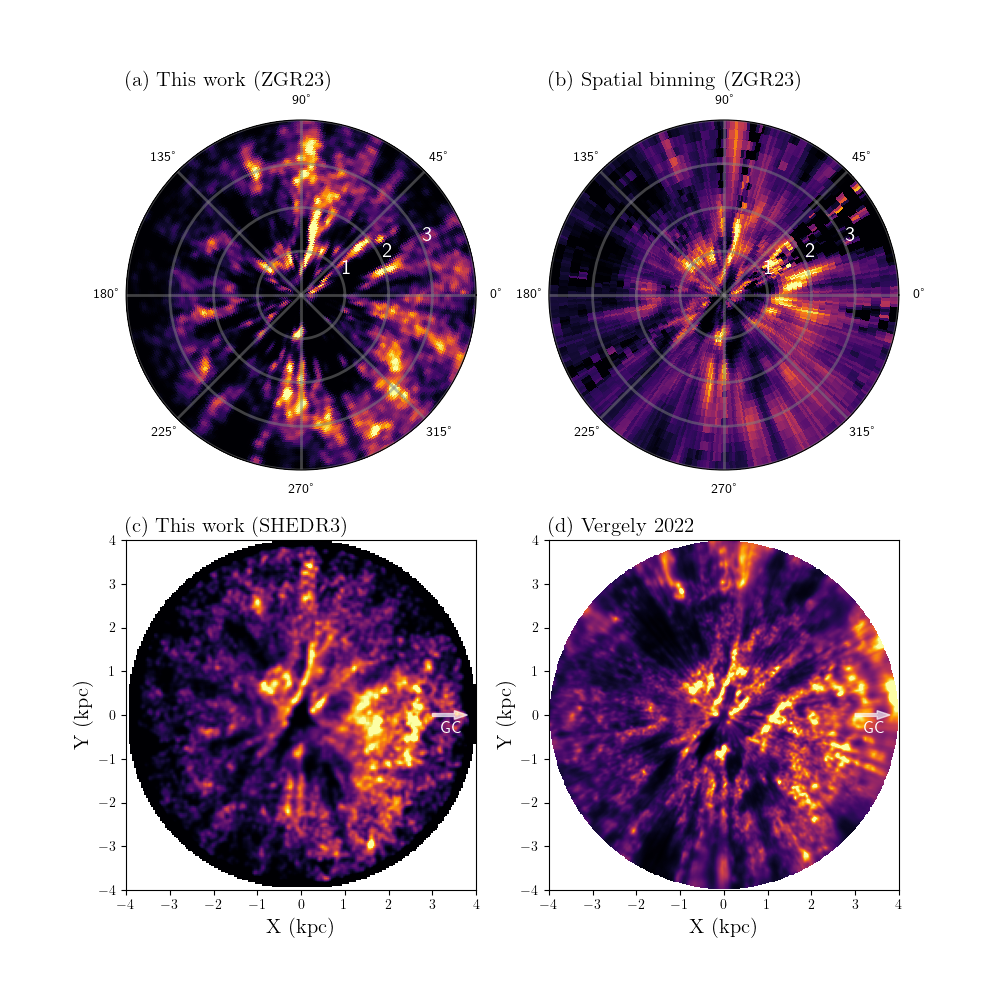}
\caption{Face-on maps from different star catalogs/techniques. Panels (a) and (b) present Polar coordinate dust maps of ZGR23. Panel (a) is produced by our method, and panel (b) is generated using a simple spatial binning technique as outlined by \cite{2023MNRAS.524.1855Z}. Panel (c) is reproduced in Cartesian coordinates from the map in Sect \ref{shemap}. Panel (d) is from \citet{2022A&A...664A.174V}, using a hierarchical-inversion technique and two merged star catalogs. \label{diffmaps}}
\end{figure*}

\subsubsection{Qualitative Comparison of Different Dust Maps} \label{zxymap}
We constructed a 3D extinction map based on the ZGR23 catalog using the same methodology described above and conducted a qualitative comparison of face-on maps derived from different methods and catalogs within the same distance range. 

Panels (a) and (b) in Figure \ref{diffmaps} both utilize the ZGR23 catalog in polar coordinates, where (a) represents our proposed method and (b) uses a simple spatial binning-based mapping technique. Notably, our method effectively suppresses finger-like artifacts and recovers structures in distant and low-extinction regions. Moreover, the reconstruction of the Perseus Arm in the second and third Galactic quadrants is particularly robust. While both methods maintain consistency in large-scale structures near the Sun, our approach demonstrates significant improvements in resolving finer details in complex environments. 

Panel (c) in Figure \ref{diffmaps} presents a Cartesian projections of Figure \ref{fig:polarmap}, while (d) display hierarchical-inversion extinction map from \citet{2022A&A...664A.174V}. Both methodologies share three key traits: they utilize model-independent approaches, rely on the Gaia EDR3 astrometric backbone, and incorporate extensive optical/infrared survey data. Their top-down views align well in the solar neighborhood and inner Galaxy, but notable discrepancies emerge in some specific features. In essence, panel (c) shows a cleaner visualization by effectively mitigating ``fingers-of-God" artifacts. Panel (d), while offering higher resolution, displays discrepancies such as the Scutum-Centaurus Arm appearing more distant and dispersed, and MCs like the one at (1.7, -3) kpc appearing farther away. Indeed, panel (c) from our work exhibits significantly greater spiral arm coherence than panel (d), especially in regions toward both the inner \citep[$1.6\pm0.2$ kpc and $2.6\pm0.2$ kpc toward the Galactic center;][]{2021A&A...653A..33N} and outer Galaxy \citep[e.g.,][]{2024ApJ...977L..35S}. 

Panels (a) and (c) in Figure \ref{diffmaps} utilize the same methodology but rely on different catalogs. Their face-on views reveal systematic differences in their extinction estimates. While panel (a) results in a more compact distribution of dust clouds with reduced granularity, panel (c) more effectively resolves faint structures and exhibits prominent continuity in spiral arm features. The most notable differences between the two panels occur within the inner Galaxy, particularly toward the Galactic center, where panel (c) displays fewer elongated patterns in the face-on map but a sharper spiral feature. 

Previous studies \citep[e.g.,][]{2019ApJ...887...93G, 2024A&A...685A..82E} have conducted comparative analyses of various 3D extinction maps. However, significant methodological differences in dust mapping techniques and extinction correction approaches hinder direct comparisons between these studies. To establish definitive conclusions, controlled analyses using identical methods and stellar catalogs are essential. While variations exist in distance coverage and spatial resolution across different maps,
these factors primarily affect small-scale structures rather than the overall large-scale morphology of the solar neighborhood. 
In summary, the CU-JADE method enables detection of MC distances, even from weak extinction jumps, while simultaneously improving 3D extinction mapping through reduction of fingers-of-God artifacts.

\subsection{Individual sky regions} \label{sect:sky regions} 
\subsubsection{Cepheus Region} \label{cep}
The MCs in the Cepheus region with different distances were roughly divided into three groups based on previous studies \citep{1989ApJ...347..231G, 1997ApJS..110...21Y}: the nearby (200 pc $\lesssim d \lesssim$ 400 pc; ${\rm -9 ~km~s}^{-1}\lesssim v \lesssim$ 9 ${\rm km~s}^{-1}$), the distant (600 pc $\lesssim d \lesssim$ 1000 pc; ${\rm -26 ~km~s}^{-1}\lesssim v \lesssim$ 2 ${\rm km~s}^{-1}$), and those in the Perseus Arm ($d \gtrsim$ 3 kpc; ${\rm -37 ~km~s}^{-1}\lesssim v \lesssim$ ${\rm -30 ~km~s}^{-1}$). The distant group might overlap with the Cepheus Flare at high Galactic latitudes ($b\sim13^{\circ}$), leading to large uncertainty and debate over the distances in this area. \citet{2019ApJ...879..125Z} also measured a group of distances at $\sim 950$ pc and $~350$ pc toward the region using a single-extinction-jump model. 

\begin{figure*}[h!]
\centering
\includegraphics[width=18cm,angle=0]{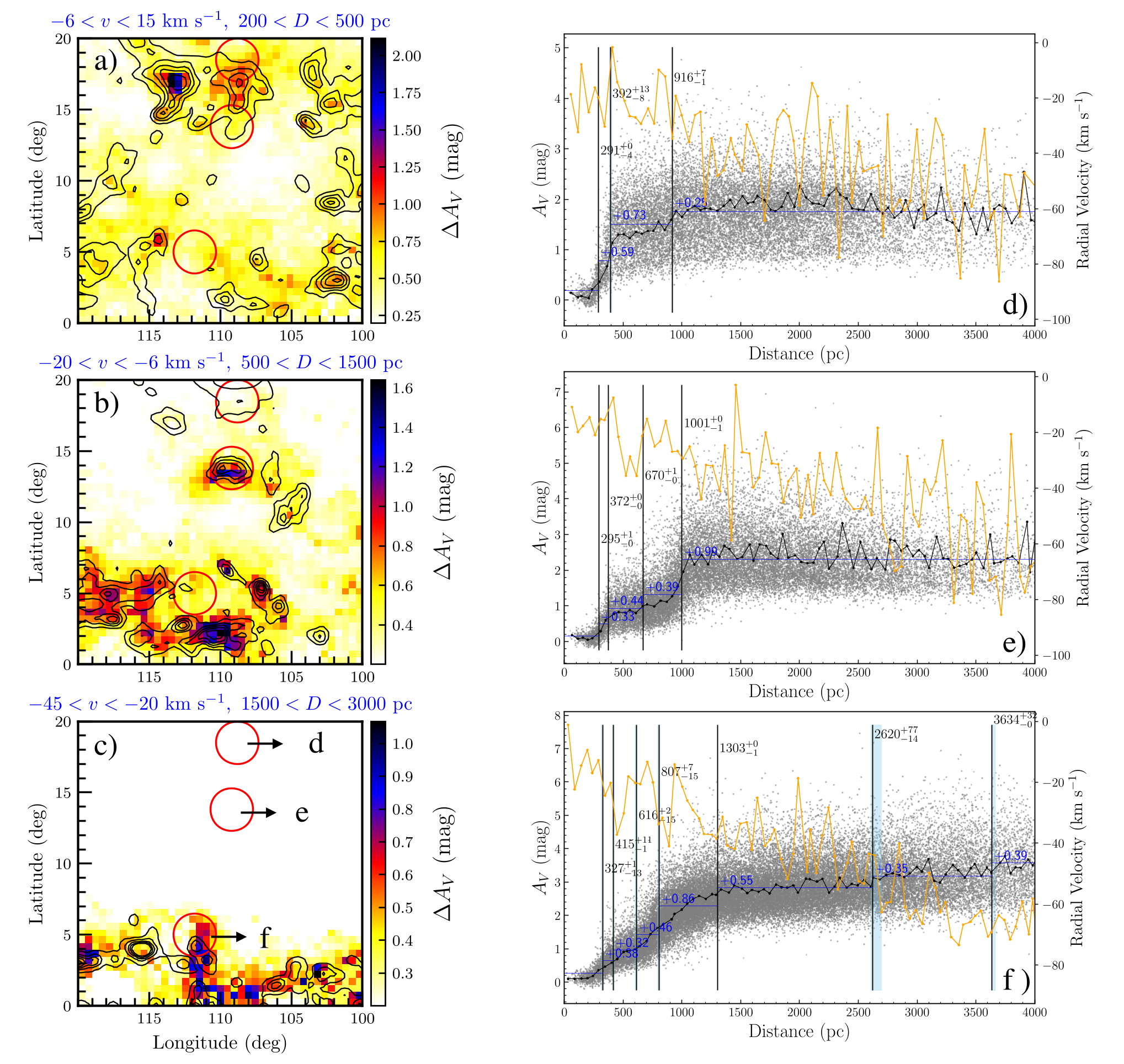}
\caption{CfA $\rm ^{12}CO$ and extinction map toward the Cepheus region. Panel (a) depicts the extinction between 200 and 500 pc using a colormap, while contours show the smoothed integrated intensity of $\rm ^{12}CO$ emission within the LSR velocity range of -6 to 15 $\rm km~s^{-1}$. Panel (b) is similar to panel (a), but focuses on the distance range in (500, 1500) pc and LSR velocities in (-20, -6) $\rm km~s^{-1}$. Panel (c) is similar to panel (a), but presents data for distances in (1500, 3000) pc and LSR velocity in (-45, -20) $\rm km~s^{-1}$. Red circles indicate regions where extinction jumps were detected in right panels (d), (e), and (f). Vertical lines represent the distances of detected extinction jumps with uncertainties in $D$-$A$ plots. Gray dots are raw stellar data points, binned every 50 pc with black lines. Blue text and horizontal lines denote the mean extinction value after accounting for extinction jumps. Orange curves are binned radial velocities of stars from the Gaia DR3. \label{fig:cepmap}}
\end{figure*}

Combined with CO data, our results give a more comprehensive view on this area. In Figure \ref{fig:cepmap}, we presented the zoom-in region between $l \in [100^{\circ}, 120^{\circ}]$ and $b \in [0^{\circ}, 20^{\circ}]$ from the all-sky map in Figure \ref{fig:edgeonmap}. Generally, MCs in Cepheus exhibit good correlations between the velocity and the corresponding distance, as well as the total gas column density and extinction (see panels (a), (b), and (c) in Figure \ref{fig:cepmap}). 
Based on the dust map in the Figure \ref{fig:cepmap} (b),
we have well determined some extended extinction structures from  high Galactic latitudes to Galactic plane (see structures at $500-1500$ pc). These extinction structures match well with the molecular gas in distant groups (black contours in panel (b) of Figure \ref{fig:cepmap}). 
This large-scale structure may extend to the junction of Cepheus and Cassiopeia (see Figure \ref{fig:edgeonmap}), where the starburst cluster M120.1+3.0 is located at $\sim 1$ kpc \citep{2023RAA....23a5019S}. 
For MCs in the most negative velocities ($ v \lesssim-20$ ${\rm km~s}^{-1}$), the corresponding extinction structures are primarily located in the Galactic plane and extend to the $b \sim 5^{\circ}$ (see circle f in panel (c) of Figure \ref{fig:cepmap}). 

In this direction, the nearby MCs at 200--500 pc (contours and colormap in panel (a) of Figure \ref{fig:cepmap}) are widely and nonuniformly distributed throughout the whole region. Studies suggest a concentration of young stars aged 1--5 Myr at distances of 330--370 pc  \citep{2021MNRAS.505.5164S}, indicating that these young stellar groups may be associated with the surrounding molecular gas at this distance.  
The nearby MCs with ${\rm -6 ~km~s}^{-1}\lesssim v \lesssim$ 15 ${\rm km~s}^{-1}$ are roughly ring-shaped, while the distant MCs (500--1500 pc; see circle e in panel (b) of Figure \ref{fig:cepmap}) are interspersed among them. These distant MCs near 1 kpc cause more significant extinction at $l \sim 109^{\circ}, b \sim 13^{\circ}$ than structures at $\sim 300$ pc (see panel (e) of Figure \ref{fig:cepmap}). 
Our results show that  there are two or even more gas layers toward the area near $l \sim 107^{\circ}-111^{\circ}$and $b \sim 13^{\circ}$ in the Cepheus Flare. 
Indeed, the Cephues Flare region exhibits multiple star-formaion episodes \citep[about 7--30 Myr;][]{2025arXiv250500407W} that appear to have significantly reshaped the gas structure and spatial distribution within 500 pc. 

The $D$-$A_{V}$ diagrams (by Mode II and $C$=2) of three directions (labeled (d), (e), and (f) in Figure \ref{fig:cepmap}) are collected from star samples within the circles. Combined with the left side-by-side plots, we further confirmed the close physical connection between CO clouds with different LSR velocities and the corresponding extinction structures in gas layers at different distances. 
At the same time, we plot the stellar radial velocities (RVs) after binning along the distance.
The RVs of the stars only roughly conforms to the trend of the rotation curve. However, we find that the dispersion and the uncertainty of the RVs of stars are very large (see orange lines in $D$-$A_{V}$ diagram of Figure \ref{fig:cepmap}). Particular caution should be taken when using stellar RV in small velocity ranges and distance intervals to study the kinematics of their structures. 
On the contrary, the RV of a young star might be useful for describing the kinematics of the young stellar objects because of their smaller kinematical ages \citep{2024ApJ...971..167Z}. 

\begin{figure*}[htbp]
\centering
\includegraphics[width=20cm,angle=0]{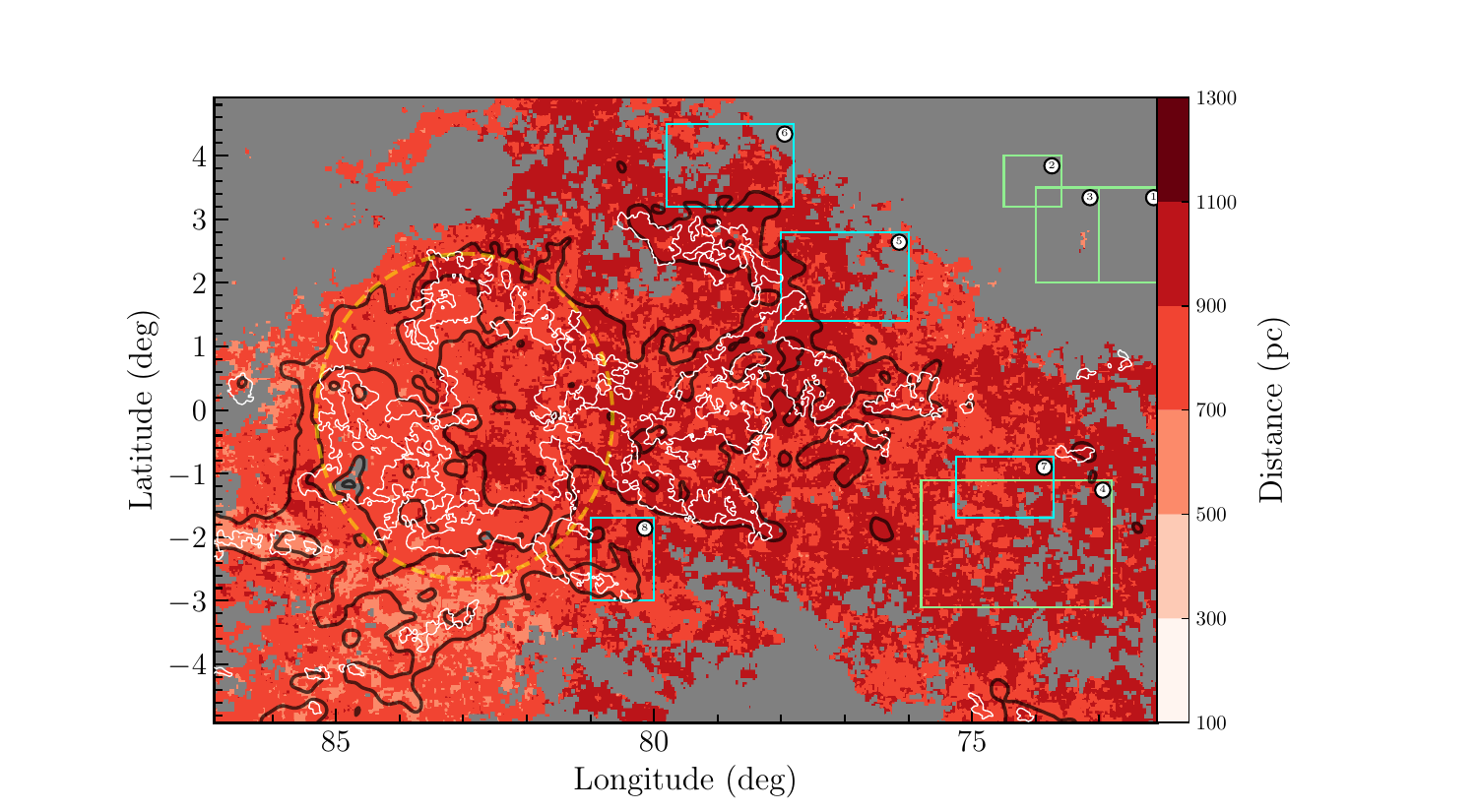}
\caption{Overlay of the identified MWISP $\rm ^{13}CO$ cloud structures (white contours) within 1.1 kpc \citep[clouds in the Cygnus Rift from][]{2024AJ....167..220Z} based on the extinction jumps detected within 1.1 kpc. The colormap shows the distance of the largest extinction in the distance interval, and the thick black contours delineate the integrated extinction in [1, 2] mag. The dashed ellipse denotes the ``800 pc loop'' identified in \citet{2024AJ....167..220Z}. The rectangles indicate areas for further distance measurements in the following. \label{fig:cygmap}}
\end{figure*}

\subsubsection{Cygnus Region} \label{cyg}
To focus on MCs near the Galactic plane, we set the sampling size $L_{0}=10'$ and sampling interval $L=2'$ for Cygnus (by Mode I and $C=1$) using $A_{G}$ of Gaia DR3. One can roughly transfer $A_{G}$ to $A_{V}$ by $A_{V}\simeq 1.35 A_{G}$ using the extinction curve from \citet{2023MNRAS.524.1855Z}. 

In our previous work \citep{2024AJ....167..220Z}, we have identified MCs in the foreground layers located on the Cygnus Rift. 
In Figure \ref{fig:cygmap}, we cross-verify the extinction structures and those $\rm ^{13}CO$ traced MCs in the foreground ($\lesssim 1.1$ kpc). Using the multijump detection method, we again confirm significant extinction in the Cygnus Rift, which is also distributed over various distance layers, e.g., structures at $\sim 1$ kpc (deep red), the 800 pc loop (light red and dashed ellipse), and closer MCs likely associated with Cyg OB7 (structures with even lighter colors below the 800 pc loop at $82^{\circ}\lesssim l \lesssim 87^{\circ}$, $-4^{\circ}\lesssim b \lesssim -2^{\circ}$). 
We can see that strong extinction structures ($\gtrsim 1$ mag, see black contours in Figure \ref{fig:cygmap}) match very well with the molecular gas traced by $\rm ^{13}CO$ (see white contours in Figure \ref{fig:cygmap}).

\begin{figure*}[htbp]
\centering
\includegraphics[width=18cm,angle=0]{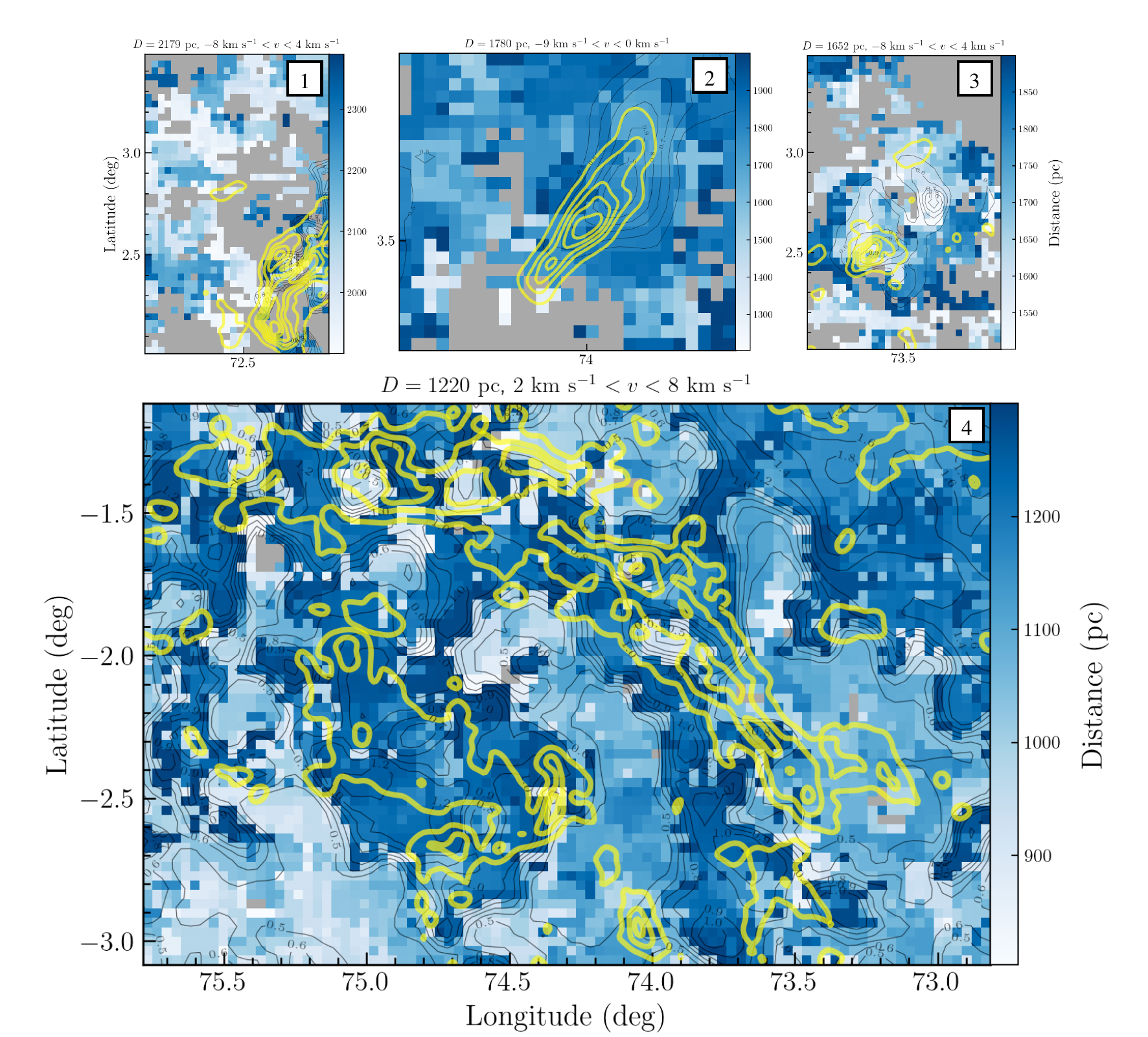}
\caption{For the case far from the Cygnus Rift, we show four examples with good matches between extinction and the integrated intensity of MWISP $\rm ^{12}CO$, and continuous distances for their structures. The colormap represents the distance of the strongest extinction component within a given distance interval along the LoS, and the colorbar shows the interval for distance measurement. Thin black contours ([0.5, 0.6, 0.7, 0.8, 0.9, 1.0] mag labeled in line) indicate the cumulative value of the extinction within this interval, outlining the structures with strong extinction in this region. Thick yellow contours represent the $\rm ^{12}CO$ emission ([5, 15, 25, 35, 45] $\rm K~km~s^{-1}$), with the integrated velocity range and the determined distance at the top of the figure. \label{fig12}}
\end{figure*}

\begin{figure*}[htbp]
\centering
\includegraphics[width=18cm,angle=0]{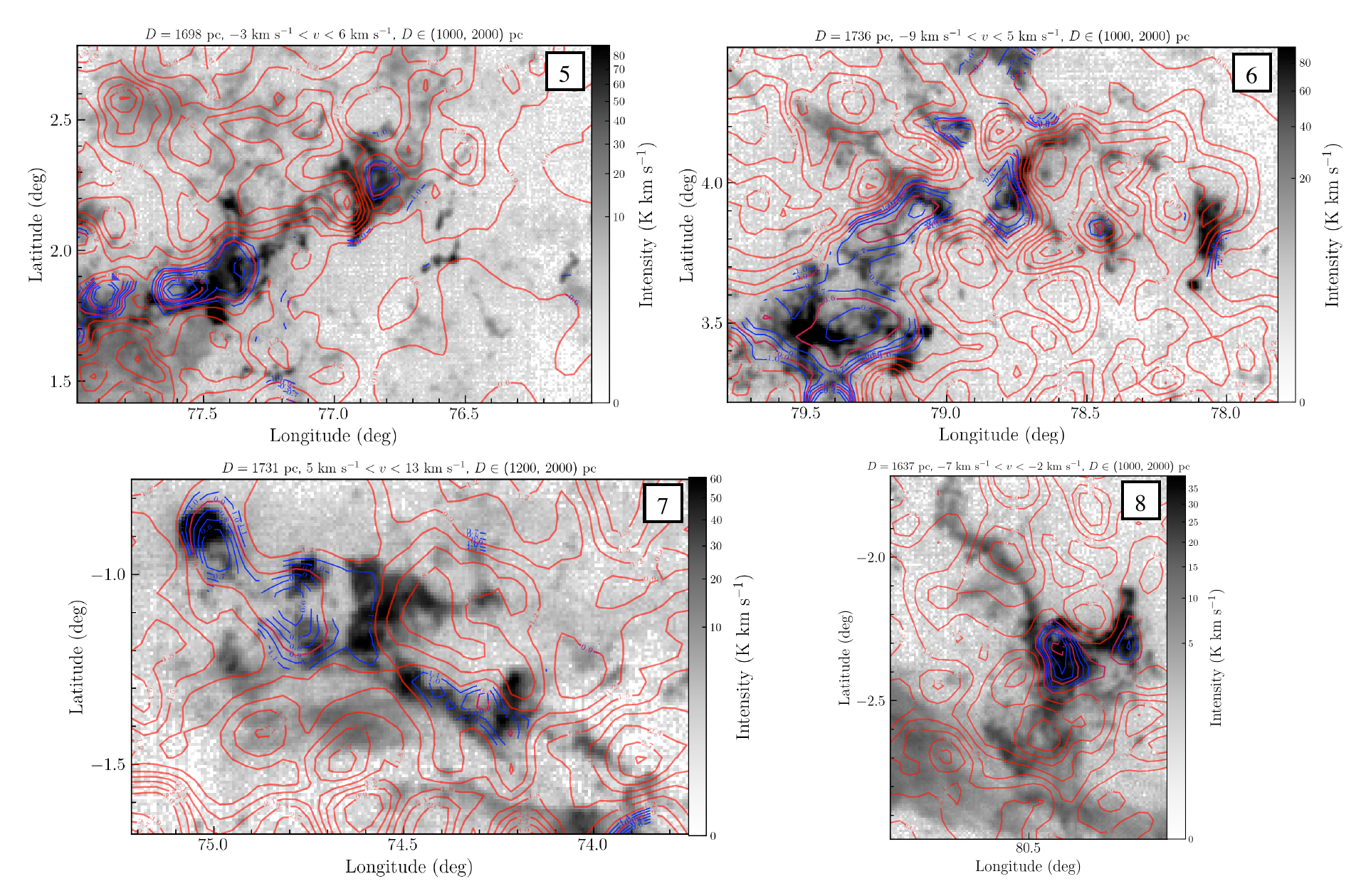}
\caption{For the case close to the projected edge of the Cygnus Rift, four examples showing an anticorrelation between extinction and integrated CO intensity in dense regions, which can also serve as strong evidence for determining distances. Here, the greyscale image represents the integrated intensity of MWISP $\rm ^{12}CO$. Similarly, thin red contours indicate the cumulative values of the extinction within the corresponding distance interval. Blue contours denote local minima of extinction (identified using $minimum\_filter$ in $Scipy.ndimage$) that coincide with dense gas regions ($\rm \gtrsim20~ K~km~s^{-1}$). The distance interval, velocity interval, and determined distances are annotated at the top of the figure. \label{fig13}}
\end{figure*}

In addition to the foreground extinction jumps, this work can provide the distances of more distant molecular gas associated with star-forming regions toward the Cygnus. 
Using our extinction maps and high-resolution $\rm ^{12}CO$ data, we successfully determine the distances to two types of MCs behind the Cygnus Rift. One type is located in areas far from the Cygnus Rift, which are relatively isolated and less affected by the extinction of the foreground gas (see the green rectangles in Figure \ref{fig:cygmap} and see Figure \ref{fig12}). 
The other type is close to the projected edge of the Cygnus Rift, where the extinction components are more complex along the LoS (see the cyan rectangles in Figure \ref{fig:cygmap} and see Figure \ref{fig13}). Excluding the unrelated gas layers in the foreground, we take more prominent extinction features (e.g., $A_{G} \gtrsim 0.6$ mag in Figure \ref{fig12}, and $A_{G} \gtrsim 1.5$ mag in Figure \ref{fig13}) to match the detailed distribution of the $\rm ^{12}CO$ emission from the MWISP and to further determine the distances. 

We found that less obscured MCs can be well matched with the corresponding extinction structures at 1--3 kpc based on the morphological correlation between the CO clouds and extinction features (see yellow contours for MWISP CO emission and colormap from extinction map in Figure \ref{fig12}). 
We highlight the coherence of the CO emission and stronger extinction features ($\gtrsim 0.5$ mag, see black contours in Figure \ref{fig12}). 

In more complex regions, the integrated intensity of the CO emission is also positively correlated with the extinction within the certain distance interval for the whole dense gas environments (see Figure \ref{fig13}). Intriguingly, in the brightest CO emission region of the four MCs, the integrated extinction in the distance measuring interval is smaller or empty. It indicates an anticorrelation between CO brightness and the extinction in the highest column density region. 
The dense part in MCs itself creates an ``extinction wall'' \citep{1993BaltA...2..171S, 2024AJ....167..220Z} that obscures background stars, preventing the detection of the extinction components. This effect could also be revealed in the $D$-$A$ map. In Figure \ref{fig13}, these regions in column density-extinction anticorrelation are highlighted with blue contours, agreeing with dense CO emission at local minima in extinction. 

In the CygX-South region, some MCs with enhanced CO emission are associated with bright mid-infrared features \citep[e.g., cloud 5 in Figure \ref{fig13}, or Clouds A and B in][]{2006A&A...458..855S, 2007A&A...474..873S}. We also identify more mid-infrared bright MCs in \citet{2024AJ....167..220Z} and the other three MCs (clouds 6, 7, and 8 in Figure \ref{fig13}) in this work.  
These MCs near photodissociation regions are suggested to be affected by the stellar winds and radiation from Cyg OB1 and OB2 \citep{2006A&A...458..855S, 2007A&A...474..873S}, featuring a head-to-tail structure pointing toward the stars association with bright CO emission. Through distance measurements with accurate Gaia parallaxes, for the first time, we confirm that these cometary MCs are located at a distance of $\sim 1.7$ kpc within an uncertainty of 10\% -- 15\%. These results strongly suggest that these interesting clouds are physically associated with the nearby HMSFRs \citep[i.e., main clusters of Cyg OB1 and OB2 at $\sim 1.7-1.8$ kpc; see][]{2019MNRAS.484.1838B, 2021MNRAS.502.6080O, 2021MNRAS.508.2370Q}. 

Unfortunately, in regions with particularly high column densities \citep[e.g., see details in Figures 8, 9, and 10 in][]{2024AJ....167..220Z}, 
star sampling in the $D$-$A$ map is incomplete due to intense extinction of multiple gas layers. It is difficult to measure the dense gas structure behind the Cygnus Rift based only on the Gaia parallax data and visible $V$/$G$-band extinction. 
Such scenarios could also be seen toward Aquila Rift region (see Figure \ref{fig:polarmap}), where the number of stars drop suddenly at a certain distance. In the future, we will combine data from the near-infrared surveys \citep[e.g.,][]{2017arXiv171103234K} and new models \citep[e.g.,][]{2018A&A...618A.168R} to explore the molecular gas distribution behind the ``extinction wall''. 



\section{summary} \label{sec6} 
We present the first application of the CUSUM-based jump point analysis method, named CU-JADE, to identify multiple extinction jumps along the LoS. Our results demonstrate CU-JADE's capability to detect weak extinction-jump points (where $\Delta A_{\lambda}\approx\sigma$) in $D$-$A$ diagrams, even in complex Galactic disk regions with overlapping gas features. The method iteratively analyzes $D$-$A$ diagrams using CUSUM statistics, focusing on cumulative extinction increments rather than traditional curve fitting for $D$-$A$ data series. Through mathematical derivation and numerical simulations, we established the relationships between $\Delta A_{\lambda}$, $\sigma$, and $CL$, while evaluating the impacts of sample sparsity (low stellar density $n$) and diffuse gas contributions (i.e., the linear slope $k$ in $D$-$A$ diagrams).

Applied to mock data, CU-JADE shows excellent accuracy and completeness in distance measurements. Validation against HMSFR maser parallaxes revealed no systematic offsets (see Figure \ref{fig3}). Using SHEDR3 data, we constructed an all-sky 3D extinction map within 4 kpc (Figures \ref{fig:edgeonmap} and \ref{fig:polarmap}), whose dense components correlate well with YOCs and masers. The map reveals known structures (e.g., Split, Radcliffe Wave, and Local Bubble clouds) while providing new insights into their 3D distribution. Comparison of our findings with the ZGR23 dataset reveals good agreement in local structures, while also identifying previously unresolved features in regions toward the outer Galaxy and the Galactic center. Notably, our results exhibit a reduced ``Finger of God" effect in top-down projections. By optimizing parameter adjustments and selecting stellar samples, CU-JADE's effectiveness and capabilities can be further enhanced.

We conducted detailed case studies on two typical regions, 
Cepheus and Cygnus, by using the CU-JADE. For Cepheus in relatively high Galactic latitude, we have reconstructed a multilayer extinction structures at different distances: the nearby MCs with $\sim 0$ $\rm km~s^{-1}$ at $\sim 300$ pc distributed throughout the whole region, intermediate-distance clouds with $\sim -12$ $\rm km~s^{-1}$ at $\sim 700-1000$ pc extend from the Galactic plane to $b\sim 15^{\circ}$, and Perseus Arm clouds at $\sim$ $2-3$ kpc (Figure \ref{fig:cepmap}). 

In the more complicated Cygnus region near the Galactic plane, CU-JADE successfully delineates both the Cygnus Rift MCs within 1 kpc (Figure \ref{fig:cygmap}) and more distant MC structures at $1-3$ kpc (Figures \ref{fig12} and \ref{fig13}), enabling the first distance measurements to MCs associated with Cyg OB2/OB1 through correlation with MWISP CO data. Being able to measure distances to MCs associated with massive stars is a big deal because OB stars are short-lived, indicating regions of recent or ongoing star formation. This correlation shows that CU-JADE can handle more challenging environments.

The CU-JADE method shows exceptional adaptability across spatial scales (a few arcmins to several tens of arcmins) and extinction environments, enabling robust MC studies at various distances.
Our 3D extinction mapping technique permits parameter optimization to align with both in CO-traced MCs and CO-dark MCs, accounting for stellar sampling density and local extinction conditions. Validation confirms the method's effectiveness for both nearby ($\lesssim 1$ kpc) and more distant ($1-3$ kpc) structures. Future work will focus on three main areas: (1) determining precise distances for MWISP CO-traced MCs using multiband data, (2) investigating structures related to CO-dark and/or CO-faint MCs, and (3) refining parameter space exploration to enhance the fidelity of maps in complex Galactic regions. This approach offers a powerful tool for linking dust extinction features with the morphological characteristics of MCs in the ISM. 

In our current work, we find that a single parameter value ``$C$" is sufficient for generating reliable thresholds across the entire sky to identify significant extinction-jump structures. This simplified approach is justified given our limited prior knowledge of dust distribution, while minimizing the model's dependence on the output. Consequently, Mode I (Section \ref{subsec:332}) represents a more conservative strategy for detecting extinction jumps, since it disregards variation in both source density and extinction properties along different sightlines (e.g., spatial fluctuations in extinction curve parameters and stellar densities). 

Future studies will focus on three key improvements to the dust map: (1) optimization of Mode II through implementation of adaptive scaling factors to correct for spatial inhomogeneities in stellar-density distributions, (2) minimization of inherent projection effects in our sampling grid using HEALPix-based resolution methods \citep{2005ApJ...622..759G}, and (3) Investigation of alternative grid classifications to recover local extinction structures (e.g., $\lesssim 500$ pc), employing a progressive analysis strategy from high Galactic latitude regions ($|b|\gtrsim 30^{\circ}$) to the Galactic disk. Additionally, independent gas tracers (e.g., HI and CO) will provide critical validation for both identifying and compensating for potential systematic biases throughout the optimization procedure. 

\begin{acknowledgments}
Acknowledgment. 
We gratefully acknowledge the referee’s expert feedback, which significantly enhanced the paper’s methodological robustness and interpretive depth. 
This research made use of the data from the Milky Way Imaging Scroll Painting (MWISP) project, which is a multiline survey in $\rm ^{12}CO$/$\rm ^{13}CO$/$\rm C^{18}O$ along the northern Galactic plane with the PMO 13.7 m telescope. We are grateful to all the members of the MWISP working group, particularly the staff members at the PMO-13.7m telescope, for their long-term support. MWISP was sponsored by the National Key R\&D Program of China with grants 2023YFA1608000, 2017YFA0402701, and the CAS Key Research Program of Frontier Sciences with grant QYZDJ-SSW-SLH047. This work is supported by NSFC grants 12173090. X.C. acknowledges the support  from the Tianchi Talent Program of Xinjiang Uygur Autonomous Region and the support by the CAS International Cooperation Program (grant No. 114332KYSB20190009). This work also made use of data from the European Space Agency (ESA) mission Gaia (https://www.cosmos.esa.int/gaia), processed by the Gaia Data Processing and Analysis Consortium (DPAC, https://www.cosmos.esa.int/web/gaia/dpac/consortium). Funding for the DPAC has been provided by national institutions, in particular the institutions participating in the Gaia Multilateral Agreement.
\end{acknowledgments}

%

\vspace{5mm}


\software{astropy \citep{2013A&A...558A..33A, 2018AJ....156..123A, 2022ApJ...935..167A}, Scipy \citep{2020NatMe..17..261V}}

The online catalogs of all-sky LoS extinction jumps, along with the CUSUM-based algorithm scripts supporting the findings of this study, are available at ScienceDB: \href{https://doi.org/10.57760/sciencedb.24650 }{10.57760/sciencedb.24650}. 



\appendix

\section{Complications of Multijump Points Detection} \label{appa} 
\subsection{The CUSUM statistics} \label{subsec:a1}

An ideal series with $j$ jump points consists of $j+1$ subsets of values, each following a Gaussian distribution (an example of mock data can be seen in Figure \ref{fig2}): 
\begin{equation}
\begin{aligned}
\\
\{A_{1}^{1}, A_{2}^{1}, \cdots, A_{i}^{1}, \cdots\}  & \sim N\left(\mu_{1}, \sigma_{1}\right), ~D \in \left[d_{min}, d_{1}\right]   \\
\{A_{1}^{2}, A_{2}^{2}, \cdots, A_{i}^{2}, \cdots\}  & \sim N\left(\mu_{2}, \sigma_{2}\right), ~D \in \left[d_{1}, d_{2}\right]   \\
 & \cdots ,   \\
\{A_{1}^{j}, A_{2}^{j}, \cdots, A_{i}^{j}, \cdots\}  & \sim N\left(\mu_{j}, \sigma_{j}\right), ~D \in \left[d_{j-1}, d_{j}\right]   \\
\{A_{1}^{j+1}, A_{2}^{j+1}, \cdots, A_{i}^{j+1}, \cdots\}  & \sim N\left(\mu_{j+1}, \sigma_{j+1}\right), ~D \in \left[d_{j}, d_{max}\right].   \\ 
\end{aligned}
\end{equation}

$D$ is the distance set in an ascending order, $A_{i}^{j}$ is the $i$th $A_{\lambda}$ value of $j$th group. $\mu_{j}$ and $\sigma_{j}$ are the mean and dispersion of $A_{\lambda}$ in $j$th group. $d_{min}$ and $d_{max}$ denote the minimum and maximum distances of the samples, respectively. Additionally, $d_{j}$ is the distance corresponding to the $j$th jump point. In our work, the mean values $\mu_{1}$, $\mu_{2}$, $\cdots$, $\mu_{j}$, $\mu_{j+1}$ are ordered such that $\mu_{1} < \mu_{2} < \cdots < \mu_{j} < \mu_{j+1}$. Similarly, for most cases, $\sigma_{1} \le \sigma_{2} \le \cdots \le \sigma_{j} \le \sigma_{j+1}$. 
As with Equation (1), the mean $A_{\lambda}$ value of all samples is 
\begin{equation}
\bar{Y} = \frac{\sum\limits^{j+1}\sum\limits^{}A_{i}^{j}}{N}. 
\end{equation}
$N$ represents the number of samples. The $D$-$S$ plot exhibits $j$ turning points, and the spike is often located at the midpoints of these turning points, 
which is initially identified as a change point (see Figure \ref{fig1}). It should be noted that the spike does not necessarily indicate the most prominent change point. However, the mean $A_{\lambda}$ values on both sides are closest to $\bar{A}$. 

The calculation of $S^{0}_{\rm diff}$ for the $D$-$S$ curve with multiple jump points is not easily derived directly from the mathematical expression. We thus simplify the problem as follows.
During iterations, it happens that a subset contains only two sets of $A_{\lambda}$ values with one extinction jump. 
The mean values of the two groups are $\mu_{j}$ and $\mu_{j+1}$, respectively, with the number of observations in each group being $N_{j}$ and $N_{j+1}$. To simplify the analysis of cumulative data, fluctuations caused by the dispersion of $A_{\lambda}$ are canceled out. The prominence of the CUSUM statistic $S$ is 
\begin{equation}
S_{\rm diff}^{0} \sim \frac{N_{j} N_{j+1}}{N_{j}+N_{j+1}}\left(\mu_{j+1}-\mu_{j}\right) \propto N ~\Delta A_{\lambda}. 
\end{equation}

In addition to those factors discussed in Section \ref{subsec:31}, the other scenarios would also affect the multiple extinction-jump detection. For example, the closer the jump point is to the middle of the sequence, the larger the $S^{0}_{\rm diff}$ value will be. The same situation also appears when the more subsequent jump points are detected (e.g., from 1 to 3 in Table \ref{tab1}). In other words, the first jump point at the middle sequence is more likely to be an $FP$. 
This scenario might affect all of our traversed jumps and can be identified using numerical methods. However, the overall impact of these factors is minor, which can be further confirmed by simulations (see Section \ref{mock} and Table \ref{tab1}).

\renewcommand\thefigure{\Alph{section}\arabic{figure}}
\setcounter{figure}{0}
\begin{figure*}[b!]
\centering
\includegraphics[width=18cm,angle=0]{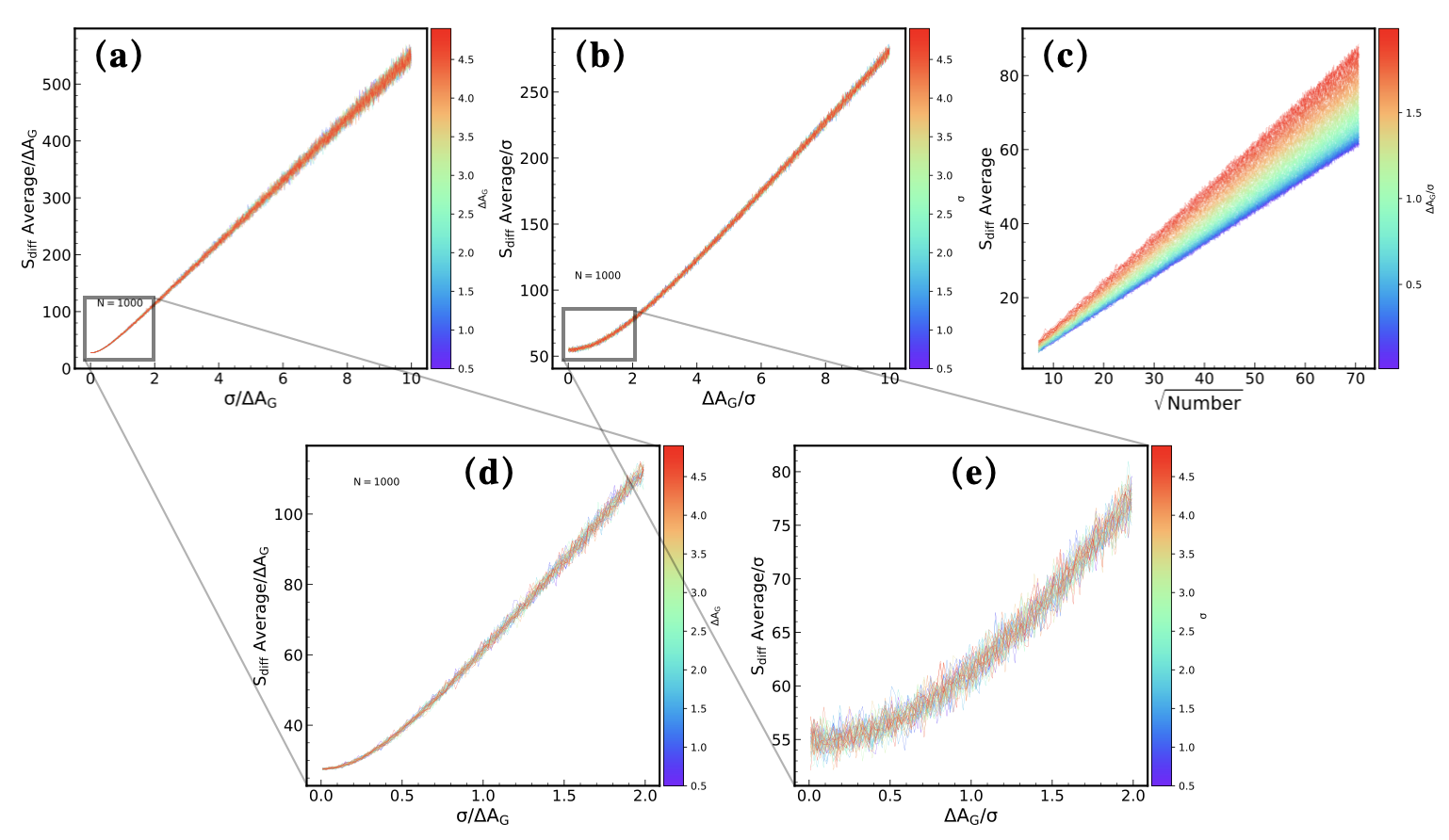}
\caption{An example for simple one-jump-point case, denoting the relationship between $\rm S_{diff}^{Avg}$, $\Delta A_{G}$, $\sigma$, and $n$. (a) Set stars number $N$=1000, $S^{\rm avg}_{\rm diff} \propto \sigma$ when $\sigma \gtrsim \Delta A$. (b) $N$=1000, $S^{\rm avg}_{\rm diff} \propto \Delta A$ when $\Delta A \gg \sigma$. (c) $S^{\rm avg}_{\rm diff} \propto \sqrt{n}$. 
The zoom-in area is presented for comparable $\Delta A_{G}$ and $\sigma$.  \label{fig:appa}}
\end{figure*}
\setcounter{figure}{0}

\subsection{The ``noise'' of CUSUM} \label{subsec:a2}

We use the statistic from a random process to serve as an estimator of the noise level for calculating $CL$, 
thereby balancing $A_{\lambda}$ and $\sigma$ based on their competitive relationship. 
Additionally, straightforward statistical measures, such as the variance of the data sequence, can also be employed. 

In both simulations and applications to real data, we have determined that the $S_{\rm diff}$, which is based on a random process, effectively assesses $CL$ for detecting jump points. 
We will now delve into a detailed discussion of the mathematical formulation of the shuffled $S_{\rm diff}$ and its influence on the outcomes. 

The expression of $S_{\rm diff}$ cannot be provided in an analytical form, and it can only be estimated by combining formula derivation and numerical computation. 
$S_{\rm diff}$ represents a set of statistical results after shuffling the samples. Because individual values lack statistical significance, we calculate their average $S^{\rm avg}_{\rm diff}$. In mathematics, the mean value $S^{\rm avg}_{\rm diff}$ is the expected value of random walk problem with variable steps. Assuming there are $j$ jump points, then $S^{\rm avg}_{\rm diff}$ is the sum of the expected values $E\left[\left|S_n\right|\right]$ of $j+1$ segments, with each segment's step length conforming to the Gaussian distribution. For one of the segments, the expected value 
\begin{equation}
E\left[\left|S_n\right|\right]=\int_{-\infty}^{\infty}|x| \frac{1}{\sqrt{2 \pi n \sigma^2}} e^{-\frac{(x-n \mu)^2}{2 n \sigma^2}} d x, 
\end{equation}
where $\mu$ is the absolute difference between the mean of each segment and the overall mean $\bar{A}$, $\sigma$ is the dispersion of $A_{\lambda}$ of each segment, and $n$ is the equivalent number of steps. 
This expression does not have a simple elementary function form, but it can be split into two parts and solved using the error function ($\rm erf$), which is a related function of the Gaussian integral, 
\begin{equation}
E\left[\left|S_n\right|\right]=\sqrt{\frac{2 n \sigma^2}{\pi}} e^{-\frac{n \mu^2}{2 \sigma^2}}+n \mu\left(1-2 \Phi\left(-\frac{n \mu}{\sqrt{n \sigma^2}}\right)\right). 
\end{equation}
Here, $\Phi$ refers to the cumulative distribution function (CDF) of the standard normal distribution, which is related to the $\rm erf$ and can be expressed as 
\begin{equation}
\Phi(z)=\frac{1}{2}\left[1+\operatorname{erf}\left(\frac{z}{\sqrt{2}}\right)\right]. 
\end{equation}

However, each step of the random walk mentioned above, which is distributed according to a Gaussian distribution, is not entirely independent. We have performed numerous simulation and calculations to verify the relationship between $N$, $\mu$, and $\sigma$, as illustrated in Figure \ref{fig:appa}. The relative magnitudes of $\mu$ and $\sigma$ significantly influence the function's expression, while $S^{\rm avg}_{\rm diff}$ is consistently proportional to the square root of the stellar number density. 

In the case of weak extinction jumps, Equation (A5) can be readily solved. For instance, when $\mu=0$, the expected value $E\left[\left|S_n\right|\right]$ equals to $\sigma \sqrt{\frac{2}{\pi}n}$. At this point, it is evident that $S^{\rm avg}_{\rm diff}$ is proportional to $\sqrt{n}$ and $\sigma$. As $\mu$ varies, the competitive relation between $\mu$ and $\sigma$ becomes evident. Their interplay are modeled by $\xi\left(\Delta A_{\lambda}, \sigma\right)$. 
In a simple one-extinction-jump case, the formula can be well fitted by 
\begin{equation}
\xi\left(\Delta A_{\lambda}, \sigma\right) \approx \frac{\Delta A_{\lambda}}{1+e^{-(\Delta A_{\lambda}-\phi\sigma)}}+\frac{\phi\sigma}{1+e^{-(\phi\sigma-\Delta A_{\lambda})}}. 
\end{equation}
Here, $\phi$ serves as a weight coefficient, which is greater than 1.

The intricate statistical relationship between $S^{0}_{\rm diff}$ and $S_{\rm diff}$ allows for more effective utilization of physical quantities such as $\Delta A_{\lambda}$, $\sigma$, and $n$, aligning closely with outcomes from manual verification. 
Moreover, validations and implementations in this work demonstrate that the derived ``noise'' in CUSUM offers an efficient approach to assess the $CL$.

\renewcommand\thefigure{\Alph{section}\arabic{figure}}
\setcounter{figure}{1}
\begin{figure}[b!]
\plotone{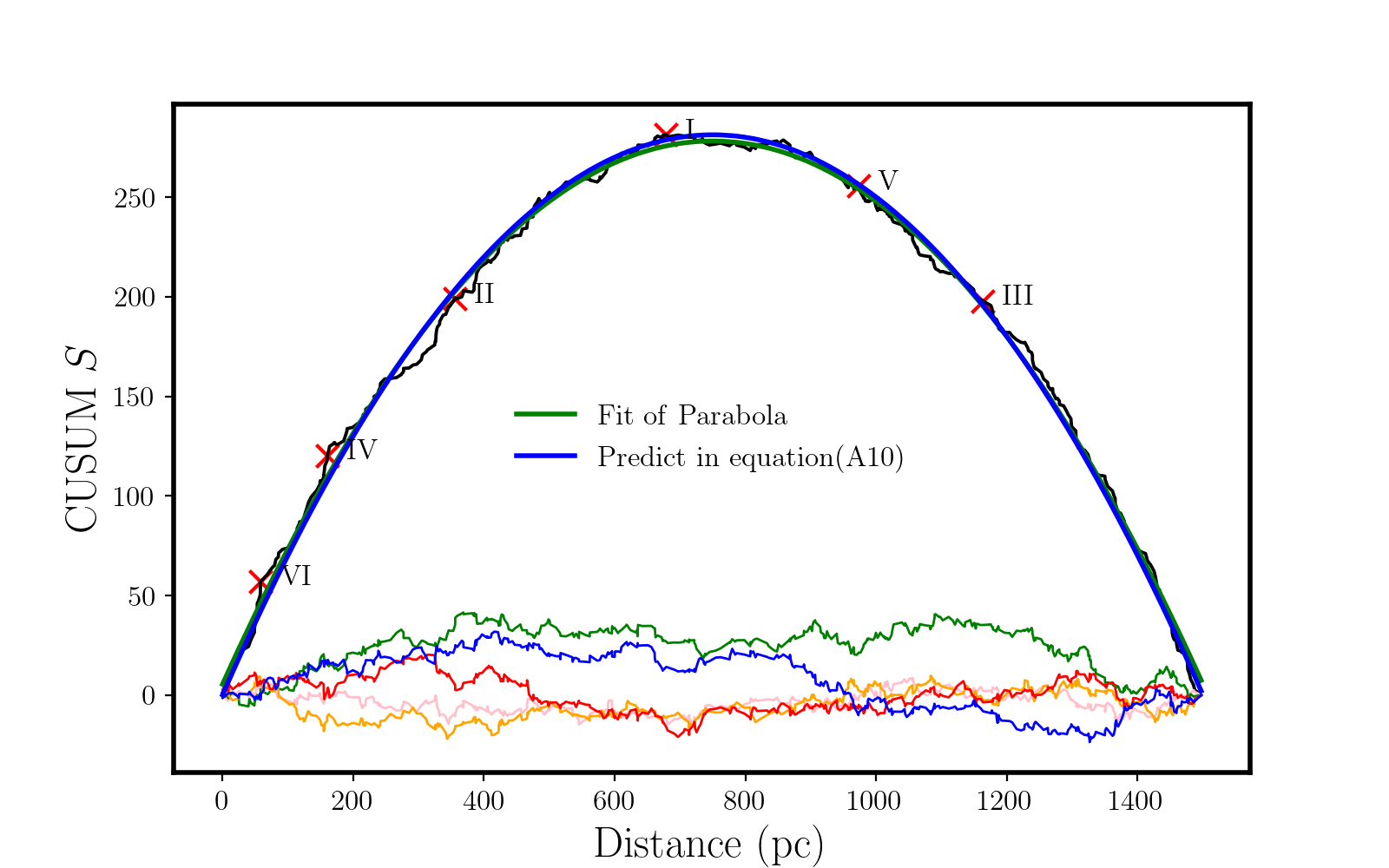}
\caption{The $D$-$S$ plot in the presence of a slope in $D$-$A$ space. Six false change points are noted by red crosses (I, II, III, IV, V, VI are the serial numbers of the detection in iterations). The curve is best fit by using the least squares method (in green). In comparison, a predicted parabola curve (in blue) is drawn using our preset parameters. The colored curves are the $D$-$S$ of shuffled samples. \label{figa2}} 
\end{figure}
\setcounter{figure}{1}

\subsection{The False Jump Features from Slope in D-A Map} \label{subsec:a3}
Obviously, some cases with small $\Delta A_{\lambda}$, small $n$, and a large $\sigma$ can result in a trade-off. This trade-off on 
$CL$ increases the likelihood of detecting weak jump points but also increases the risk of containing false detections. 
However, the LoS cumulative extinction of the diffuse gas over a large distance range could also cause false detection. 
In the following analysis, we assume that the density of the diffuse gas exhibits a roughly uniform distribution with distance. 

We assume the extinction of diffuse gas still follows a Gaussian distribution on the $D$-$A$ map, yet it is superimposed on a linear function. The CUSUM statistic $S$ satisfies 
\begin{equation}
S\left(x+\frac{1}{n}\right)-S\left(x\right)=\bar{Y}-kx-m, ~x>0, ~k>0, ~m\ge0, 
\end{equation}
where $x$ is the distance of each star, $n$ is the number density of stars, $k$ is the slope, and $m$ is the intercept, which is equal to the $A_{\lambda}$ in the foreground. $S\left(x\right)$ is a quadratic function plus a periodic function. In our case, the periodicity of this function comes from the unresolved distance interval of stars, so here we propose that the quadratic function $S\left(x\right)$ follows: 
\begin{equation}
S\left(x\right)=-\frac{1}{2}nkx^{2}+\left[n\left(\bar{Y}-m\right)+\frac{1}{2}k\right]x+const. 
\end{equation}
In the case, we set $S\left(0\right)=0$. The location of the symmetry axis, and/or the first false change point to be detected, is in $x_{\rm I}=\frac{\left(\bar{Y}-m\right)}{k}+\frac{1}{2n}$. Because $\frac{1}{2n}$ is smaller than the distance resolution in a given number density $n$, we ignore it and $x_{\rm I}\approx \frac{\left(\bar{Y}-m\right)}{k}$. CUSUM $S$ can thus be written as
\begin{equation}
S\left(x\right)=-\frac{1}{2}nkx^{2}+n\left(\bar{Y}-m\right)x. 
\end{equation}
The prominence $S_{\rm diff}$ is, 
\begin{equation}
S_{\rm diff}^{0}=\frac{n}{2k}\left(\bar{Y}-m\right)^{2}=\frac{1}{2}nk{x_{\rm I}}^{2}. 
\end{equation}
Once the distance range is fixed, $S_{\rm diff}^{0}$ is proportional to number density $n$ and slope $k$.

As illustrated in Figure \ref{figa2}, the $D$-$S$ plot draws a parabolic trace as expected. In a smaller $CL$ threshold ($C$=1 in Mode I), six false change points were detected in order, all of which are located near the midpoints of their respective iteration intervals. 
In comparison, we give the best fit by using the least squares method (green curve in Figure \ref{figa2}) and a predicted parabola curve from the preset parameters (blue curve in Figure \ref{figa2}) by Equation (A10). A stricter criterion is needed to avoid such false detections. Next, we have to further analyze the 
properties of ``noise'' $S_{\rm diff}$.

\renewcommand\thefigure{\Alph{section}\arabic{figure}}
\setcounter{figure}{2}
\begin{figure}[b!]
\plotone{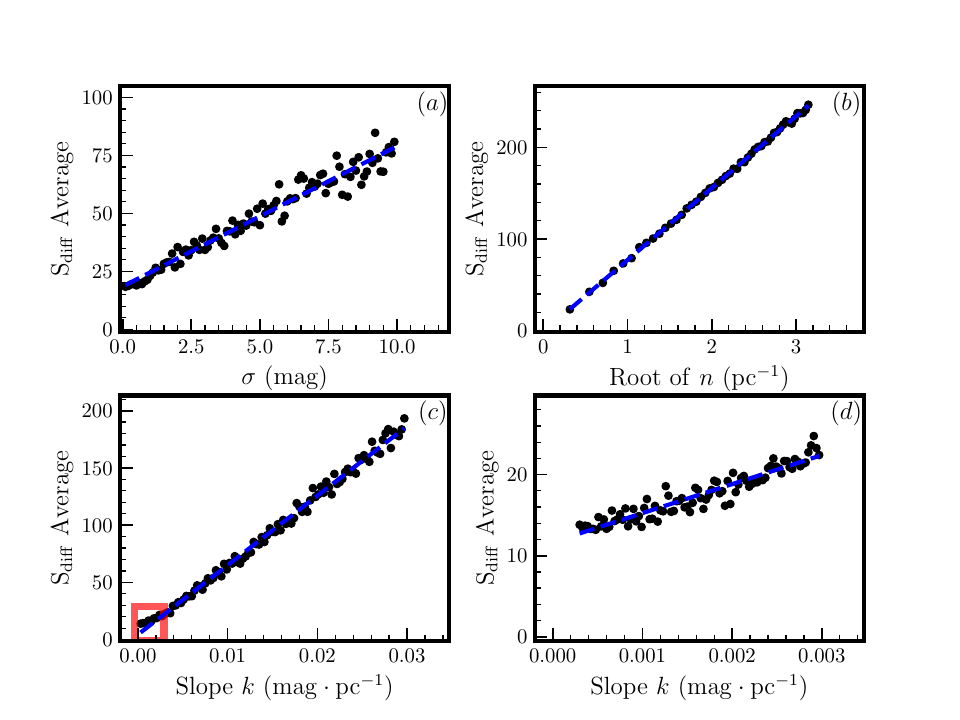}
\centering
\caption{Relations between $S_{\rm diff}$ and $A_{\lambda}$ dispersion $\sigma$, number density $n$, and slope $k$. (a) Set $n=0.1$ ${\rm pc}^{-1}$ and $k=0.003$ ${\rm mag}\cdot{\rm pc}^{-1}$, $S^{\rm avg}_{\rm diff} \propto \sigma$ when $\sigma$ becomes larger. (b) $S^{\rm avg}_{\rm diff} \propto \sqrt{n}$. (c) Set $\sigma$=1 mag, the relation between $S^{\rm avg}_{\rm diff}$ and $k$ shows a similar trend to panel (a). Panel (d) is a zoom-in from panel (c). 
\label{figa3}}
\end{figure}
\setcounter{figure}{2}

Similar to the situation where there are jump points, the $S_{\rm diff}$ with a slope of $k$ and dispersion $\sigma$ follows a random walk with variable steps (see Appendix \ref{subsec:a1}). 
Providing an analytical form for $S_{\rm diff}$ is equally complex. Generally, $S_{\rm diff}$ is proportional to $\sqrt{n}$, while $k$ and $\sigma$ also present a competitive relationship. 
For example, when $k$ is small, $\sigma$ becomes dominant. The relation between $S_{\rm diff}$, $n$, $k$, and $\sigma$ can also be derived using mock data. 
By examining the typical values from the $D$-$S$ diagram along the LoS, we can describe the relationship between $S_{\rm diff}$ and one of the quantities.

In Figure \ref{figa3} (a), a linear correlation between $S_{\rm diff}$ average and $\sigma$ exists with a intercept term $\sim 20$ when we use a constant $n=0.1$ ${\rm pc}^{-1}$ and $k=0.003$ ${\rm mag}\cdot{\rm pc}^{-1}$. 
When fixing up the $\sigma$ (1 mag) in Figure \ref{figa3} (c), $S_{\rm diff}$ is roughly proportional to $k$. 
When $k\gtrsim 0.003$ ${\rm mag}\cdot{\rm pc}^{-1}$, both $S_{\rm diff}$ and $S_{\rm diff}^{0}$ are proportional to $k$. Therefore the $CL$ of false detection is independent of slope $k$. This presents a difficulty in identifying true $\Delta A_{\lambda}$, leading to the false detection from the large $k$.  

However, we find that $\sigma$ domains $S_{\rm diff}$ in small $k$ values. Figure \ref{figa3} (d) shows a zoomed-in view of the small $k$ region, where $S_{\rm diff}$ climbs slowly within the range of small $k$. In a simple test of the mock data, assuming the following preset parameters:
a typical value of jumps intervals $x_{\rm I}=750$ pc, a gradient $k=0.002$ ${\rm mag}\cdot{\rm pc}^{-1}$, a stars density $n=0.1~{\rm pc}^{-1}$, and $A_{\lambda}$ dispersion $\sigma=1~{\rm mag}$, we compute the prominence of the preset slope. 
The calculated prominence is $S_{\rm diff}^{0}\sim 50$ and the averaged $S_{\rm diff}\sim 20$. This can effectively avoid false detections caused by slopes when setting $C\geq 2$, and the percentile to 0.95. 
Of course, for larger $\Delta A_{\lambda}$, we could raise the confidence threshold to ensure a more reliable result. 
On the other hand, if $k\gtrsim 0.003$ ${\rm mag}\cdot{\rm pc}^{-1}$, we can find these change points near midpoint of each iteration, due to the symmetric/parabolic nature of $S_{\rm diff}$. This may help us determine if a extinction structure is from a real jump or a large slope $k$ (see Figure \ref{figb1} in Appendix \ref{app:appb}). 

\section{Performance on complex mixed LoS components} \label{app:appb}

The chosen region likely consists of two or more sets of dust components along the LoS. Consequently, the distribution of $A_{\lambda}$ versus distance is not a simple staircase but a complex superposition of overlapping components. This occurs when unknown components or an off-cloud region are included, necessitating an examination of whether our method remains effective. The $A_{\lambda}$ values of stars between jump points are described as following a multi-Gaussian distribution, which poses challenges for modeling with analytical solutions. In this case, we generate mock data consisting of two groups of $A_{\lambda}$ values with positive $\Delta A_{\lambda}$ increments. The two group of mock data have different preset jump points.  

\renewcommand\thefigure{\Alph{section}\arabic{figure}}
\setcounter{figure}{0}
\begin{figure*} [hb!]
\gridline{\fig{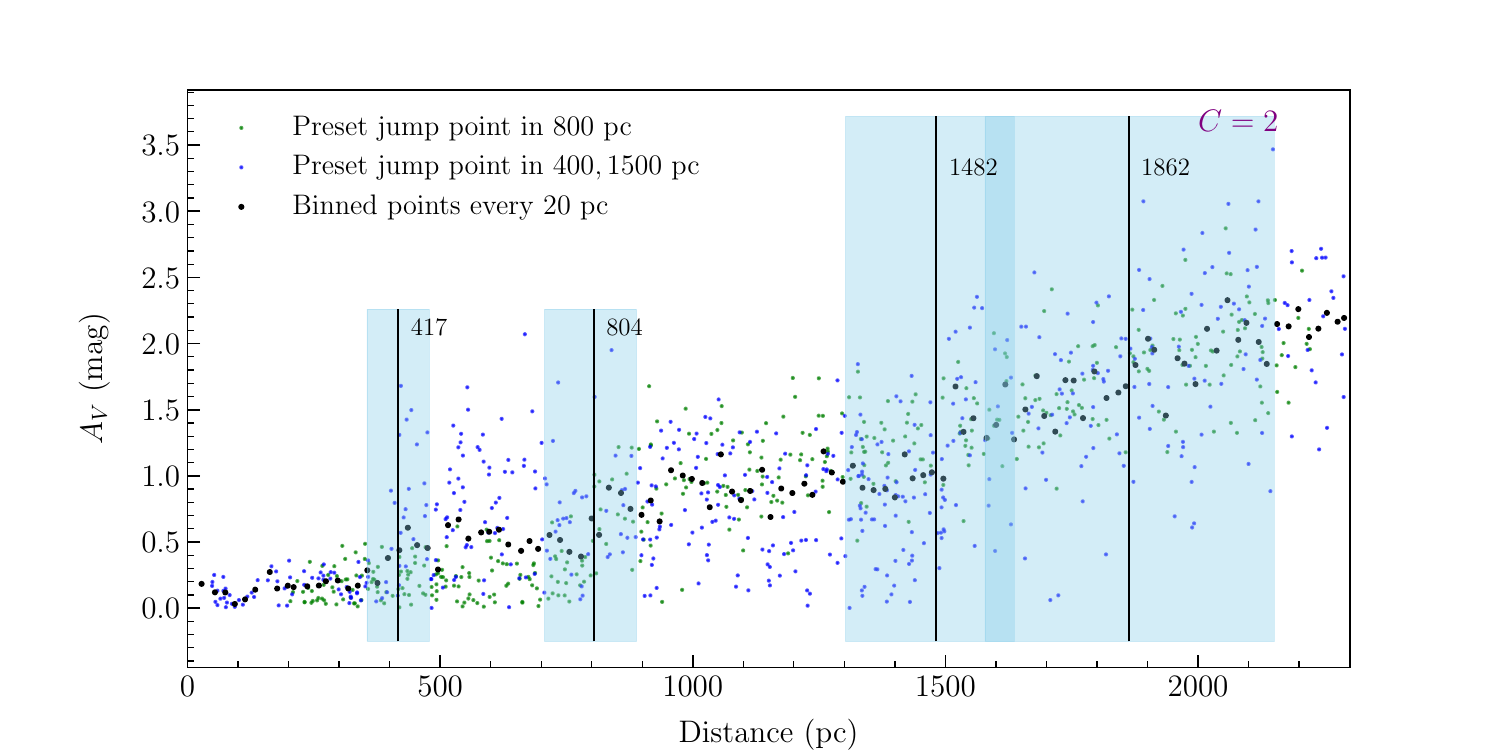}{0.54\textwidth}{(a)}
    \hspace{-11mm}
\fig{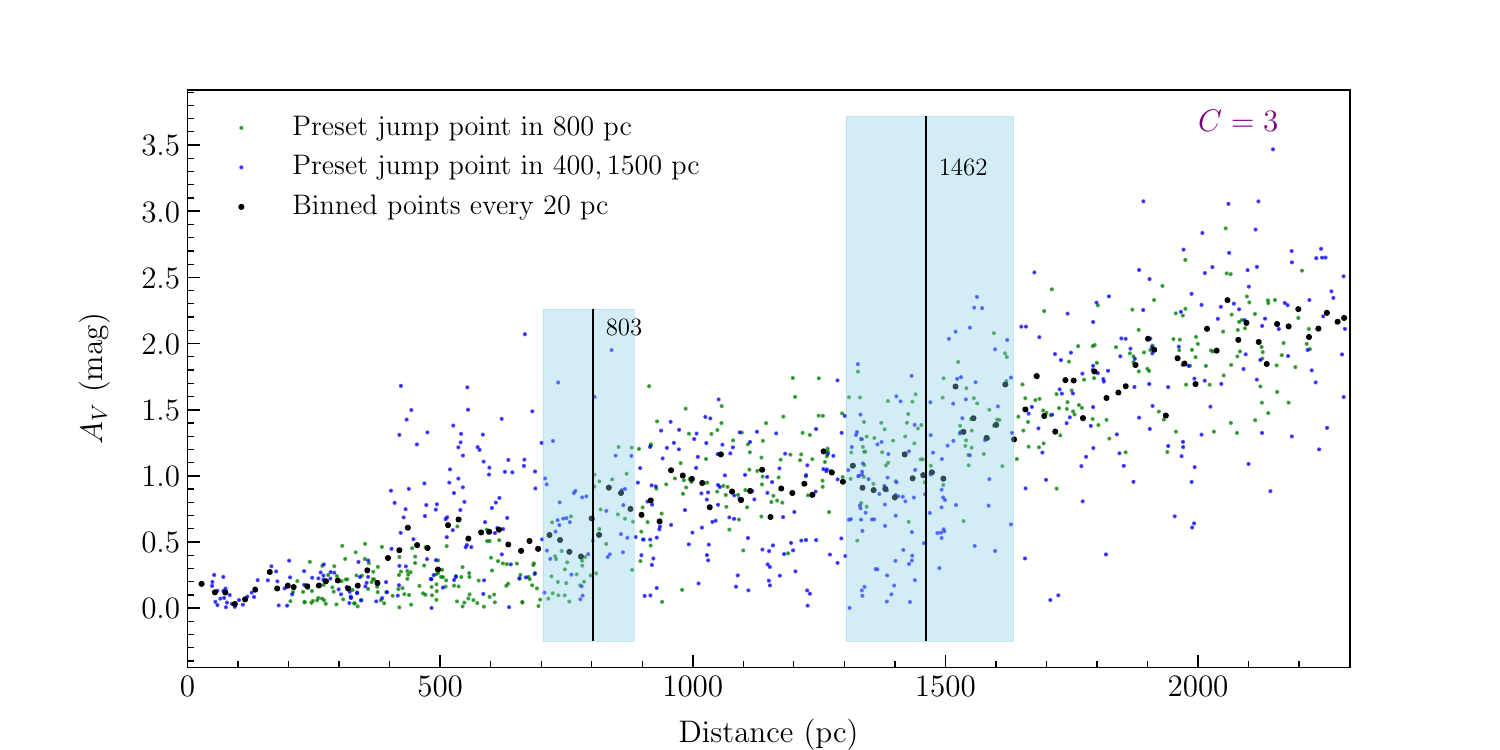}{0.54\textwidth}{(b)} }
\caption{The blue dots are preset to indicate jump points at two locations, 400 pc and 1500 pc, while the green dots are preset to indicate a jump point at 800 pc. In panel (a), $C$ is set to 2, and in panel (b), $C$ is set to 3, with 4 and 2 jump  points detected, respectively. 
\label{figb1}}
\end{figure*}
\setcounter{figure}{0}

In Figure \ref{figb1}, blue dots represent jump points preset at 400 pc and 1500 pc, respectively. Green dots indicate a single jump point preset at 800 pc. The number density of stars ranges from $0.2~{\rm pc}^{-1}$ to $0.5~{\rm pc}^{-1}$, which is a summation of 0.2 and 0.3 ${\rm pc}^{-1}$. The slope is set to 0.001 $\rm mag~pc^{-1}$ in the distance range $1.5 \sim 2.3$ kpc, while the slopes of the remaining distance intervals can be disregarded. Those $\Delta A_{\lambda}$ are set to be $\sim0.5$ mag. 

When we preset the number density levels at $n\sim0.5~{\rm pc}^{-1}$, and $\Delta A_{\lambda} \sim 0.5~{\rm mag}$, the constant 
$C=3$ is quite stringent to identify the extinction jumps. As shown in panel (b) of Figure \ref{figb1}, our model identifies two out of three jump points with confidence intervals that encompass the prior values. In panel (a), when $C$ is reduced to 2, all jump points are detected, along with a false detection near the midpoint of the $1.5 \sim 2.3$ kpc, which is caused by the previously set slopes (as discussed in Appendix \ref{subsec:a3}).

\section{Examples for HMSFRs associated with masers} \label{app:appc}

\renewcommand\thefigure{\Alph{section}\arabic{figure}}
\setcounter{figure}{0}
\begin{figure*}[h!]

\gridline{\fig{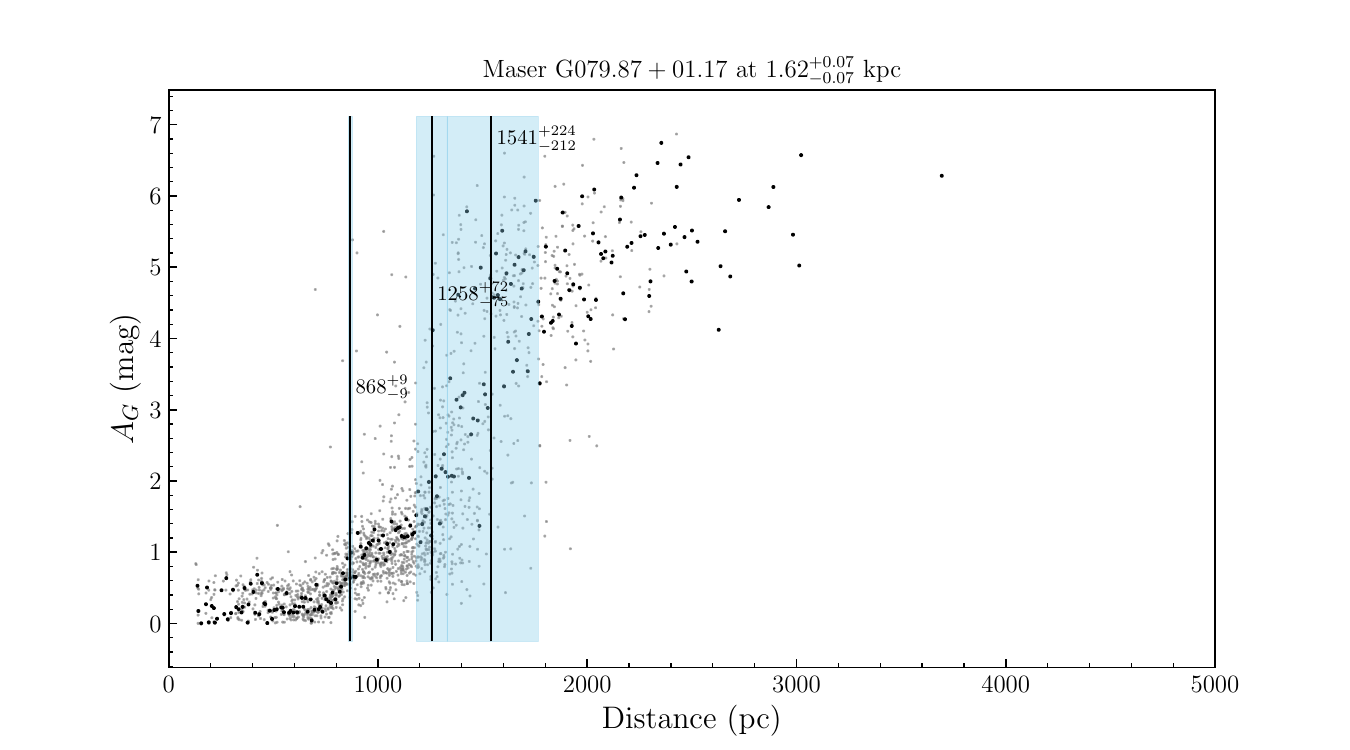}{0.5\textwidth}{(a)}
    \hspace{-15mm}
\fig{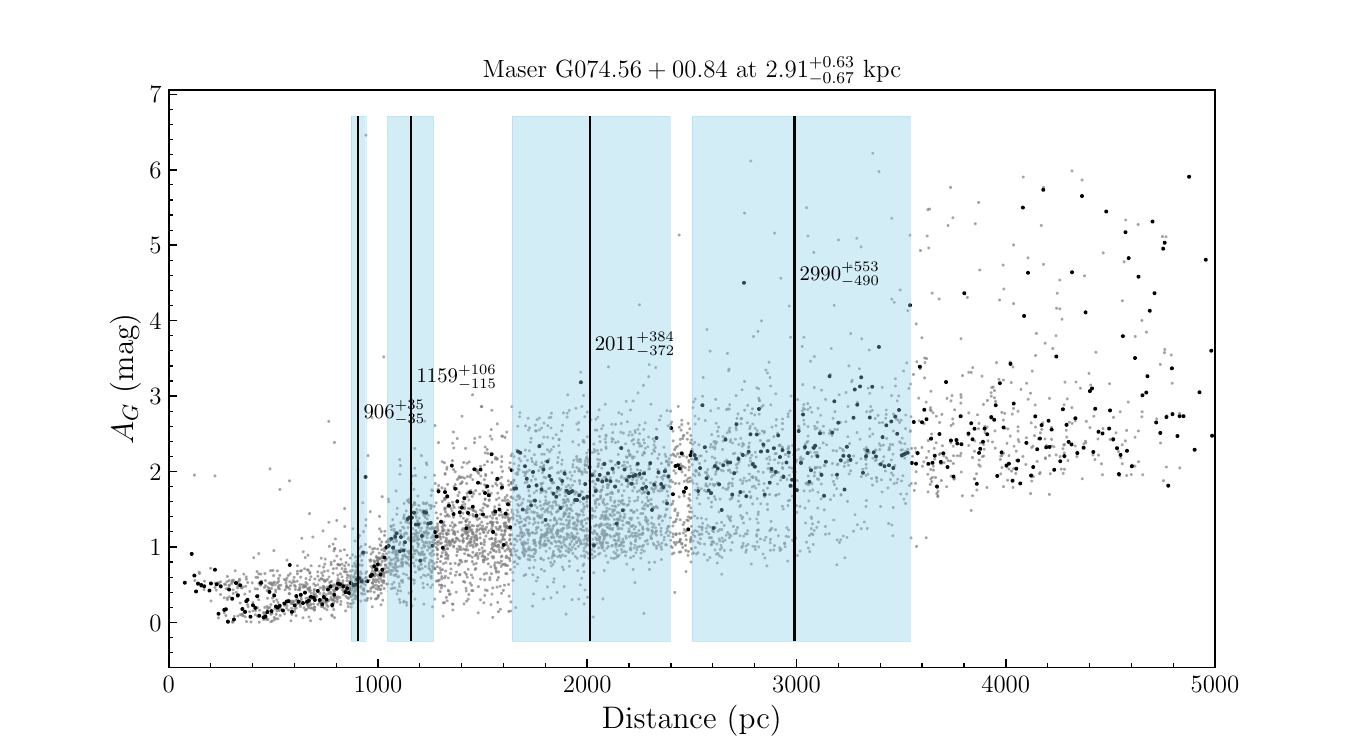}{0.5\textwidth}{(b)}
          }
\vspace{-5mm}
\gridline{\fig{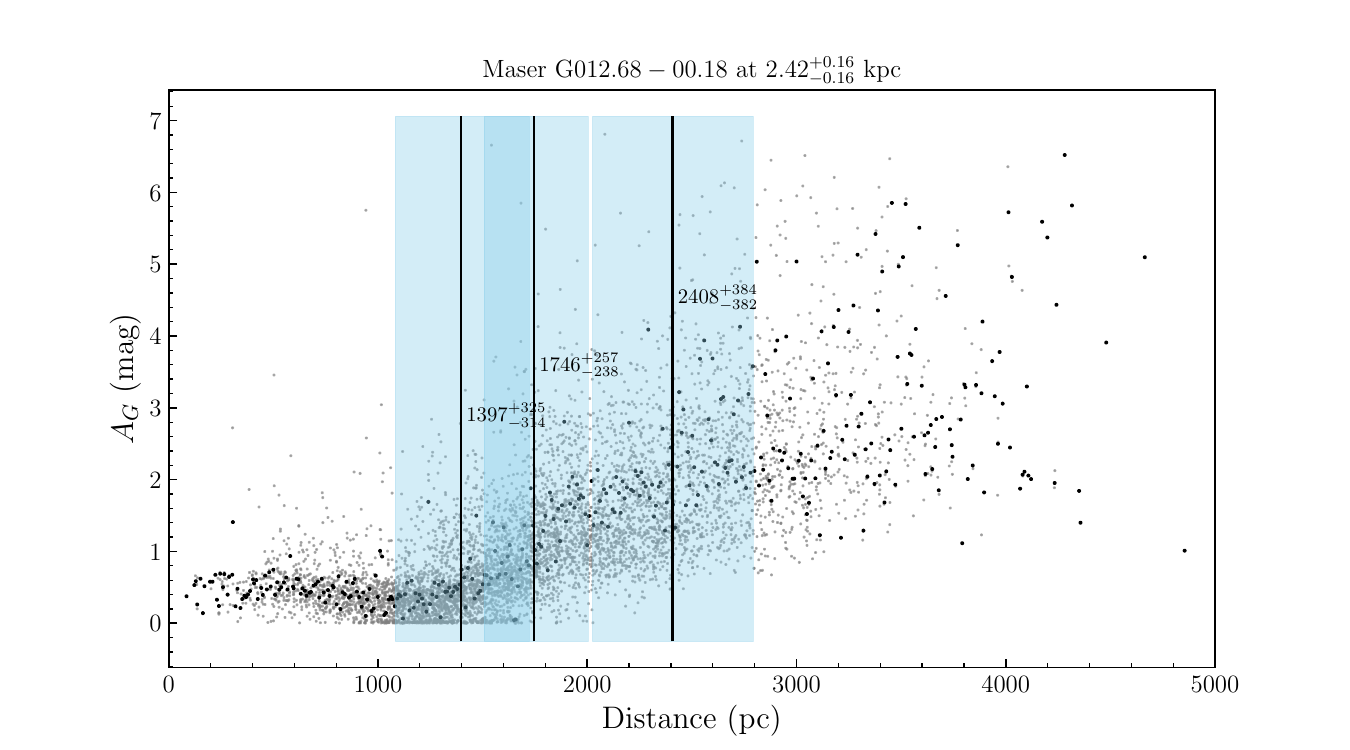}{0.5\textwidth}{(c)}
    \hspace{-15mm}
\fig{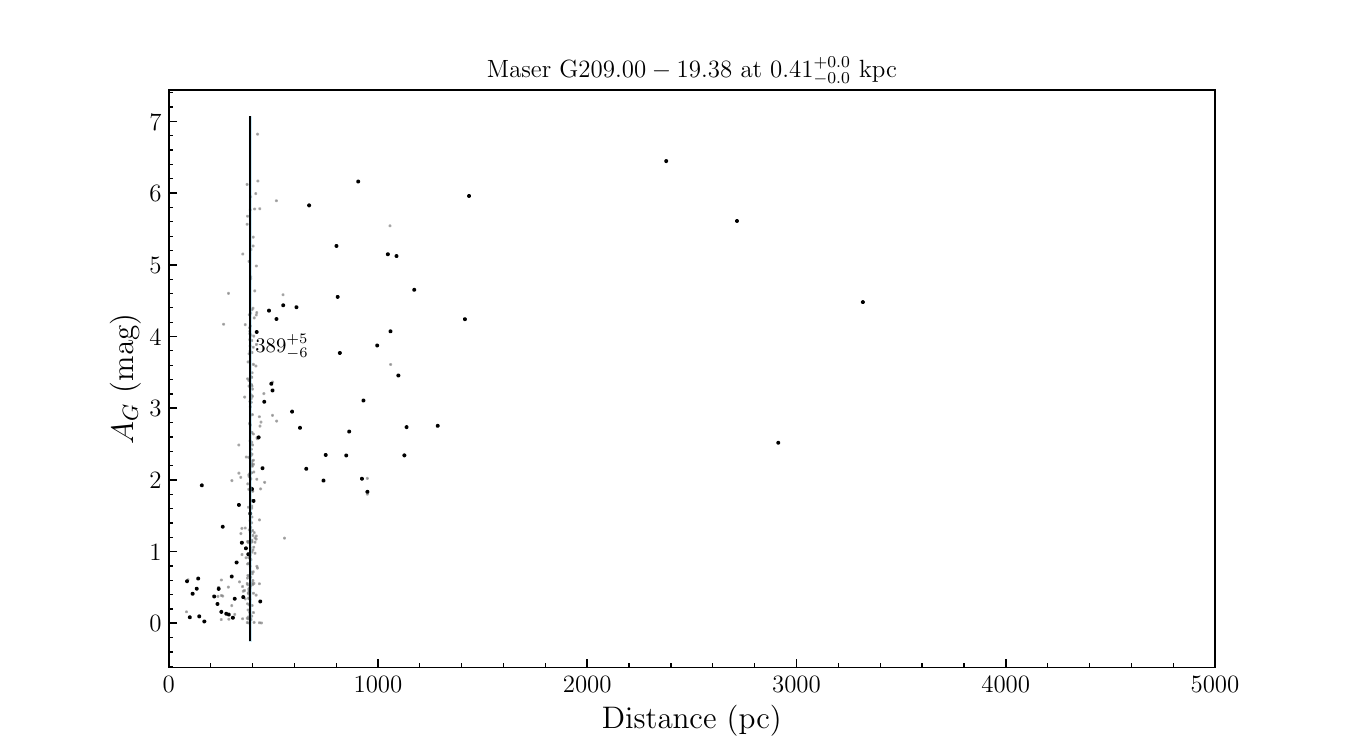}{0.5\textwidth}{(d)}
          }
\caption{Four examples for G079.87+01.17, G074.56+00.84, G012.68-00.18, and G209.00-19.38. Stars are represented by gray dots, while black dots denote binned points every 20 pc. All four examples corresponding to the masers are located at the last jump point. 
\label{figc1}}
\end{figure*}
\setcounter{figure}{0}

In Section \ref{validatemaser}, we found that a total of 46 masers align well with our results. Here, we provide four examples (G079.87+01.17,  G074.56+00.84, G012.68-00.18, and G209.00-19.38) of $D$-$A$ diagrams (see Figure \ref{figc1}) toward the direction of masers. The distance of the last jump point agrees well with the masers. 



\bibliography{multi_jump}{}
\bibliographystyle{aasjournal}



\end{document}